\batchmode
\makeatletter
\def\input@path{{D:/Dropbox/0_ClimatePrediction/Paper3_Gyrostat/Paper1_conservative/arXiv_v2/}}
\makeatother
\documentclass[11pt,english]{article}
\usepackage{lmodern}

\let\origrmdefault\rmdefault
\usepackage[math]{iwona}
\renewcommand{\rmdefault}{\origrmdefault}
\usepackage[T1]{fontenc}
\usepackage[latin9]{inputenc}
\usepackage{geometry}
\usepackage{babel}
\usepackage{mathrsfs}
\usepackage{amsmath}
\usepackage{amssymb}
\usepackage{graphicx}
\usepackage{setspace}
\usepackage[pdftex,unicode=true,pdfusetitle,
 bookmarks=true,bookmarksnumbered=false,bookmarksopen=false,
 breaklinks=false,pdfborder={0 0 1},backref=false,colorlinks=false]
 {hyperref}

\makeatletter

\newcommand*\LyXZeroWidthSpace{\hspace{0pt}}
\providecommand{\tabularnewline}{\\}

\newcommand{\lyxaddress}[1]{
	\par {\raggedright #1
	\vspace{1.4em}
	\noindent\par}
}

\@ifundefined{date}{}{\date{}}
\usepackage{pdflscape}

\makeatother

\begin{document}
\title{Invariants and chaos in the Volterra gyrostat without energy conservation\thanks{This paper is dedicated to the memory of Professor A S Vasudeva Murthy.}}
\author{Ashwin K Seshadri\textsuperscript{1} and S Lakshmivarahan\textsuperscript{2}}
\maketitle

\lyxaddress{\textsuperscript{1}Centre for Atmospheric and Oceanic Sciences and
Divecha Centre for Climate Change, Indian Institute of Science, Bangalore
560012, India. Email: ashwins@iisc.ac.in.}

\lyxaddress{\textsuperscript{2}Emeritus faculty at the School of Computer Science,
University of Oklahoma, Norman, OK 73012, USA. Email: varahan@ou.edu.}

\subsection*{Declarations of interest: none}

\pagebreak{}
\begin{abstract}
The model of the Volterra gyrostat (VG) has not only played an important
role in rigid body dynamics but also served as the foundation of low-order
models of many naturally occurring systems. It is well known that
VG possesses two invariants, or constants of motion, corresponding
to kinetic energy and squared angular momentum, giving oscillatory
solutions to its equations of motion. Nine distinct subclasses of
the VG have been identified, two of which the Euler gyroscope and
Lorenz gyrostat are each known to have two constants. This paper characterizes
quadratic invariants of the VG and each of its subclasses, showing
how these enjoy two invariants even when rendered in terms of a non-invertible
transformation of parameters, leading to a transformed Volterra gyrostat
(TVG). If the quadratic coefficients of the TVG sum to zero, as they
do for the VG, the system conserves energy. In all of these cases,
the flows preserve volume. However, physical models where the quadratic
coefficients do not sum to zero are ubiquitous, and characterization
of invariants and the resulting dynamics for this more general class
of models with volume conservation but without energy conservation
is lacking. This paper provides the first such characterization for
each of the subclasses of the VG in the absence of energy conservation,
showing how the number of invariants depends on the number of linear
feedback terms. It is shown that the gyrostat with three linear feedback
terms has no invariants. The number of invariants circumscribes the
possible dynamics for these three-dimensional flows, and those without
any invariants are shown to admit rich dynamics including chaos. This
gives rise to a broad class of three-dimensional volume conserving
chaotic flows, arising naturally from model reduction techniques.
\end{abstract}

\section*{Keywords}

Volterra gyrostat; Low-order models; Volume conserving flows; Quadratic
invariants; Low-dimensional chaos; Dissipationless limit

\pagebreak{}

\section{Introduction}

Let $\mathrm{x}\left(t\right)=\left(x_{1}\left(t\right),x_{2}\left(t\right),x_{3}\left(t\right)\right)^{T}\in\mathbb{R}^{3}$
be the state of a dynamical system at time $t\geq0$. Let $\mathrm{f}:\mathbb{R}^{3}\rightarrow\mathbb{R}^{3}$
with $\mathrm{f}\left(\mathrm{x}\right)=\left(f_{1}\left(t\right),f_{2}\left(t\right),f_{3}\left(t\right)\right)^{T}\in\mathbb{R}^{3}$
denote the vector field of the associated dynamics where $f_{i}\left(\mathrm{x}\right)=\alpha_{i}x_{j}x_{k}$
for $i\neq j\neq k$ and $\alpha_{i}\in\mathbb{R}$ for $1\leq i\leq3$
satisfy $\alpha_{1}+\alpha_{2}+\alpha_{3}=0$. Let $\mathrm{A}$ denote
a skew-symmetric matrix of order $3$ having the form 
\begin{equation}
\mathrm{A}=\left[\begin{array}{ccc}
0 & -h_{3} & h_{2}\\
h_{3} & 0 & -h_{1}\\
-h_{2} & h_{1} & 0
\end{array}\right].\label{eq:p1}
\end{equation}
Then 
\begin{equation}
\mathrm{\dot{x}}\left(t\right)=\mathrm{f\left(x\right)}+\mathrm{Ax}\label{eq:p2}
\end{equation}
describes evolution of the state $\mathrm{x}\left(t\right)$ of a
dynamical system known as the Volterra gyrostat (VG) \cite{Volterra1899}.
When the matrix $\mathrm{A=0}$ in Eq. (\ref{eq:p2}), the resulting
dynamics is called the Euler gyroscope (EG). In the parlance of rigid
body dynamics, VG has played a central role in design and control
of mechanical systems \cite{Wittenburg1977}. More recently it was
demonstrated that VG and many of its special cases occur naturally
as the basic building blocks to additively construct a variety of
low-order, finite dimensional models for many naturally occurring
phenomena including the now famous equations of Rayleigh Benard convection
\cite{Lorenz1963,Gluhovsky2002}, vorticity dynamics \cite{Lorenz1960,Charney1979,Swart1988},
among others. Refer to \cite{Gluhovsky1999} (hereafter GT (1999))
and \cite{Gluhovsky1997,Gluhovsky2002,Gluhovsky2006}. Many atmospheric
low-order models developed through model reduction techniques \cite{Kwasniok1996,Kwasniok2007}
possess a structure that can be naturally represented as combinations
of coupled VGs. The investigation of dynamics in the presence of gyrostatic
forces and the resulting integrals of motion is an active area of
study \cite{Amer2012,Amer2021,He2022}.

It is well known that VG in Eq. (\ref{eq:p2}) enjoys two invariants:
kinetic energy and square of the angular momentum, quadratic invariants
that each restrict its trajectories to two manifolds whose intersection
is a one-dimensional manifold containing the solution trajectories.
By specializing the choice of parameters $\alpha_{i}$ and $h_{i}$,
$1\leq i\leq3$ in Eq. (\ref{eq:p2}), the authors in GT (1999) identify
nine distinct subclasses of VG, which includes the Euler gyroscope
(EG), the so-called Lorenz gyrostat (LG), and other nonlinear oscillators.
Further extensions of VG that include nonlinear feedback, called generalized
VG, are contained in the two papers \cite{Lakshmivarahan2008,Lakshmivarahan2008a}.
For an interesting survey and historical review of the gyrostat and
its applications refer to \cite{Tong2009}.

While it is known (GT (1999)) that among the cases of VG, both the
EG and LG enjoy two quadratic invariants, it is not known if the other
special cases of VG also possess two invariants. In this brief study,
using a common framework, we derive expressions for the two invariants
for all the nine special cases of VG. The existence of two quadratic
invariants is closely tied to the constraint that the $\alpha_{i}$'s
sum to zero, which originates in their physical interpretation in
terms of the principal moments of inertia of the gyrostat. Furthermore,
this constraint plays an important role in energy conservation \cite{Lakshmivarahan2008a}.

It is of interest that there is a close connection between EG and
the maximum simplification equations \cite{Lorenz1960,Lakshmivarahan2006}.
The systems are not identical, the EG obeys the constraint $\alpha_{1}+\alpha_{2}+\alpha_{3}=0$,
whereas the maximum simplification equations do not. Yet the maximum
simplification equations are also known to possess two quadratic invariants.
It is not known whether the other special cases of the VG, upon relaxing
this constraint, also enjoy two invariants, which in general could
differ from those of the VG. Such a property would give rise to simple
oscillatory dynamics, whereas the absence of any invariants is a necessary
condition for chaotic behavior in these three-dimensional models.

Prior work has considered the chaotic dynamics resulting from time-varying
perturbations to the EG (e.g., \cite{Holmes1983}). The motivation
of the present paper is to analyze the specific models represented
by Eqs. (\ref{eq:p1})-(\ref{eq:p2}), where departures from the EG
are state feedback that naturally arise in low-order models from the
effects of forcing. A comprehensive analysis for various special cases
of Eq. (\ref{eq:p1})-(\ref{eq:p2}), specializing for various subclasses
of the VG \cite{Gluhovsky1999}, is absent.

The approach of this paper allows us to consider the invariants of
the VG and its generalization and examine the role played by this
constraint. We are thus able to demonstrate that the non-existence
of any invariants in these three dimensional flows yields richer dynamics
including chaos. Section 2 contains the derivation of the quadratic
invariants of the VG, with and without energy conservation. The VG
is rendered in a modified form, first described by GT (1999), through
a non-invertible transformation of parameters. For each of the subclasses
of the resulting system, quadratic invariants are estimated by a common
procedure. We consider the effects of removing the constraint $\alpha_{1}+\alpha_{2}+\alpha_{3}=0$,
not only for the maximum simplification equations, but also the eight
other subclasses. It is shown that not all of these subclasses possesses
two quadratic invariants, giving rise to richer dynamics when this
constraint is absent. In the absence of energy conservation, the number
of invariants is closely tied to the number of linear feedback terms.
Section 3 shows that chaos can arise in those subclasses which do
not have any invariants. Thus, we identify new classes of volume conserving
chaotic flows that naturally emerge in low-order models without any
forcing or dissipation.

\section{Constants of motion for the Volterra gyrostat (VG)}

\subsection{Volterra equations for the gyrostat}

Consider Volterra's equations 
\begin{align}
K_{1}^{2}\dot{y}_{1} & =\left(K_{2}^{2}-K_{3}^{2}\right)y_{2}y_{3}+h_{2}y_{3}-h_{3}y_{2}\nonumber \\
K_{2}^{2}\dot{y}_{2} & =\left(K_{3}^{2}-K_{1}^{2}\right)y_{3}y_{1}+h_{3}y_{1}-h_{1}y_{3}\nonumber \\
K_{3}^{2}\dot{y}_{3} & =\left(K_{1}^{2}-K_{2}^{2}\right)y_{1}y_{2}+h_{1}y_{2}-h_{2}y_{1}\label{eq:p3}
\end{align}
where for $i=1,2,3$, $y_{i}$ denotes the components of the angular
velocity of the carrier body, $K_{i}^{2}=I_{i}$ are the principal
moments of inertia of the gyrostat, and $h_{i}$ are the components
of the fixed angular momentum of the rotor relative to the carrier.
The dots indicate rate of change, for e.g., $\dot{y}_{1}=dy_{1}/dt$.
Together these constitute the gyrostat, which possesses two quadratic
invariants, the kinetic energy
\begin{equation}
E=\frac{1}{2}\sum_{i=1}^{3}K_{i}^{2}y_{i}^{2}\label{eq:p4}
\end{equation}
and the magnitude of the angular momentum vector, or equivalently
one-half of its square 
\begin{equation}
M=\frac{1}{2}\sum_{i=1}^{3}\left(K_{i}^{2}y_{i}+h_{i}\right)^{2}\label{eq:p5}
\end{equation}
both of which are constant in time for Eq. (\ref{eq:p3}), i.e., $\dot{E}=\dot{M}=0$
(GT (1999)).

\subsection{Transformed Volterra gyrostat (TVG)}

Consider transformation of variables $K_{i}y_{i}=x_{i}$, introduced
by GT (1999). In state variables $x_{i}$, the gyrostat equations
are
\begin{align}
K_{1}K_{2}K_{3}\dot{x}_{1} & =\left(K_{2}^{2}-K_{3}^{2}\right)x_{2}x_{3}+K_{2}h_{2}x_{3}-K_{3}h_{3}x_{2}\nonumber \\
K_{1}K_{2}K_{3}\dot{x}_{2} & =\left(K_{3}^{2}-K_{1}^{2}\right)x_{3}x_{1}+K_{3}h_{3}x_{1}-K_{1}h_{1}x_{3}\nonumber \\
K_{1}K_{2}K_{3}\dot{x}_{3} & =\left(K_{1}^{2}-K_{2}^{2}\right)x_{1}x_{2}+K_{1}h_{1}x_{2}-K_{2}h_{2}x_{1}\label{eq:p6}
\end{align}
and, defining a new time-variable $t=K_{1}K_{2}K_{3}s$, so that
\begin{equation}
\frac{d}{ds}=\frac{d}{dt}\frac{dt}{ds}=K_{1}K_{2}K_{3}\frac{d}{dt}
\end{equation}
we obtain
\begin{align}
x_{1}' & =\left(K_{2}^{2}-K_{3}^{2}\right)x_{2}x_{3}+K_{2}h_{2}x_{3}-K_{3}h_{3}x_{2}\nonumber \\
x_{2}' & =\left(K_{3}^{2}-K_{1}^{2}\right)x_{3}x_{1}+K_{3}h_{3}x_{1}-K_{1}h_{1}x_{3}\nonumber \\
x_{3}' & =\left(K_{1}^{2}-K_{2}^{2}\right)x_{1}x_{2}+K_{1}h_{1}x_{2}-K_{2}h_{2}x_{1}
\end{align}
where $'$ denotes $d/ds$. This model satisfies $M'=dM/ds=0$. Defining
new parameters 
\begin{align}
p & =K_{2}^{2}-K_{3}^{2},q=K_{3}^{2}-K_{1}^{2},\textrm{and }r=K_{1}^{2}-K_{2}^{2}\nonumber \\
a & =K_{1}h_{1},b=K_{2}h_{2},\textrm{and }c=K_{3}h_{3}\label{eq:p9}
\end{align}
with $p+q+r=0$, the model is rendered as
\begin{align}
x_{1}' & =px_{2}x_{3}+bx_{3}-cx_{2}\nonumber \\
x_{2}' & =qx_{3}x_{1}+cx_{1}-ax_{3}\nonumber \\
x_{3}' & =rx_{1}x_{2}+ax_{2}-bx_{1},\label{eq:p10}
\end{align}
 which we denote as the transformed Volterra gyrostat (TVG). The quadratic
invariants of this model will be considered in the remainder of this
section. The symbols used in the paper are summarized in Table 4.

\subsection{Constants of motion of TVG}

Since the transformation in Eq. (\ref{eq:p6}) is smooth, we expect
TVG to possess as many constants of motion (``invariants'') as VG,
as is easily shown. A system 
\begin{equation}
\mathrm{\dot{y}}=\mathrm{g}\left(\mathrm{y}\right)\label{eq:p11}
\end{equation}
with $\mathrm{y}=\left[\begin{array}{cccc}
y_{1} & y_{2} & \ldots & y_{n}\end{array}\right]^{T}\in\mathbb{R}^{n}$ being the state vector and $\mathrm{g}=\left[\begin{array}{cccc}
g_{1} & g_{2} & \ldots & g_{n}\end{array}\right]^{T}:\mathbb{R}^{n}\rightarrow\mathbb{R}^{n}$ the vector field has $k$ distinct invariants $C_{i}\left(y_{1},y_{2},\ldots y_{n}\right)$
if
\begin{equation}
\frac{dC_{i}}{dt}=\sum_{j=1}^{n}\frac{\partial C_{i}}{\partial y_{j}}\dot{y}_{j}=0\label{eq:p12}
\end{equation}
for $i=1,2,\ldots k$. Now, if there is a smooth change of variables
$\mathrm{y=h\left(x\right)}$, where $\mathrm{h}=\left[\begin{array}{cccc}
h_{1} & h_{2} & \ldots & h_{n}\end{array}\right]^{T}:\mathbb{R}^{n}\rightarrow\mathbb{R}^{n}$ we have 
\begin{equation}
\dot{y}_{j}=\sum_{l=1}^{n}\frac{\partial h_{j}}{\partial x_{l}}\dot{x}_{l}\label{eq:p13}
\end{equation}
and substituting in Eq. (\ref{eq:p12})
\begin{equation}
\frac{dC_{i}}{dt}=\sum_{l=1}^{n}\left(\sum_{j=1}^{n}\frac{\partial C_{i}}{\partial y_{j}}\frac{\partial h_{j}}{\partial x_{l}}\right)\dot{x}_{l}=0\label{eq:p14}
\end{equation}
for $i=1,2,\ldots k$, such that any smooth change of variables must
preserve the invariants. Thus, for kinetic energy 
\begin{equation}
E=\frac{1}{2}\sum_{i=1}^{3}K_{i}^{2}\left(\frac{x_{i}}{K_{i}}\right)^{2}=\frac{1}{2}\sum_{i=1}^{3}x_{i}^{2}
\end{equation}
its evolution
\begin{equation}
E'=\sum x_{i}x_{i}'=0
\end{equation}
from Eq. (\ref{eq:p10}), since $p+q+r=0$. Similarly, for the angular
momentum
\begin{equation}
M=\frac{1}{2}\sum_{i=1}^{3}\left(K_{i}x_{i}+h_{i}\right)^{2}
\end{equation}
we obtain
\begin{equation}
M'=\sum_{i=1}^{3}\left(K_{i}x_{i}+h_{i}\right)K_{i}x_{i}'
\end{equation}
and, upon substituting from Eq. (\ref{eq:p10})
\begin{multline}
M'=x_{1}x_{2}x_{3}\left(K_{1}^{2}p+K_{2}^{2}q+K_{3}^{2}r\right)+cx_{1}x_{2}\left(r+K_{2}^{2}-K_{1}^{2}\right)+ax_{2}x_{3}\left(p+K_{3}^{2}-K_{2}^{2}\right)+bx_{1}x_{3}\left(q+K_{1}^{2}-K_{3}^{2}\right)\label{eq:p19}
\end{multline}
and, using the relations in Eq. (\ref{eq:p9}), we obtain also
\begin{equation}
M'=0.\label{eq:p20}
\end{equation}
There is an incongruity in the above analysis leading to Eq. (\ref{eq:p20}),
with Eq. (\ref{eq:p19}) containing parameters of the VG as well as
the TVG. In general, we know the representation of the model in either
one set of variables, the VG or the TVG, but not both. This would
not be a cause for difficulty if, given the TVG in Eq. (\ref{eq:p10}),
we were to to solve for the parameters including $K_{i}$, $i=1,2,3$
of the VG. However, this parameter transformation is not invertible,
as shown below. Consider Jacobian
\begin{equation}
\mathrm{H}\equiv\left[\begin{array}{cccccc}
\frac{\partial p}{\partial K_{1}} & \frac{\partial p}{\partial K_{2}} & \frac{\partial p}{\partial K_{3}} & \frac{\partial p}{\partial h_{1}} & \frac{\partial p}{\partial h_{2}} & \frac{\partial p}{\partial h_{3}}\\
\frac{\partial q}{\partial K_{1}} & \frac{\partial q}{\partial K_{2}} & \frac{\partial q}{\partial K_{3}} & \frac{\partial q}{\partial h_{1}} & \frac{\partial q}{\partial h_{2}} & \frac{\partial q}{\partial h_{3}}\\
\frac{\partial r}{\partial K_{1}} & \frac{\partial r}{\partial K_{2}} & \frac{\partial r}{\partial K_{3}} & \frac{\partial r}{\partial h_{1}} & \frac{\partial r}{\partial h_{2}} & \frac{\partial r}{\partial h_{3}}\\
\frac{\partial a}{\partial K_{1}} & \frac{\partial a}{\partial K_{2}} & \frac{\partial a}{\partial K_{3}} & \frac{\partial a}{\partial h_{1}} & \frac{\partial a}{\partial h_{2}} & \frac{\partial a}{\partial h_{3}}\\
\frac{\partial b}{\partial K_{1}} & \frac{\partial b}{\partial K_{2}} & \frac{\partial b}{\partial K_{3}} & \frac{\partial b}{\partial h_{1}} & \frac{\partial b}{\partial h_{2}} & \frac{\partial b}{\partial h_{3}}\\
\frac{\partial c}{\partial K_{1}} & \frac{\partial c}{\partial K_{2}} & \frac{\partial c}{\partial K_{3}} & \frac{\partial c}{\partial h_{1}} & \frac{\partial c}{\partial h_{2}} & \frac{\partial c}{\partial h_{3}}
\end{array}\right]
\end{equation}
which is
\begin{equation}
\mathrm{H}=\left[\begin{array}{cccccc}
0 & 2K_{2} & -2K_{3} & 0 & 0 & 0\\
-2K_{1} & 0 & 2K_{3} & 0 & 0 & 0\\
2K_{1} & -2K_{2} & 0 & 0 & 0 & 0\\
h_{1} & 0 & 0 & K_{1} & 0 & 0\\
0 & h_{2} & 0 & 0 & K_{2} & 0\\
0 & 0 & h_{3} & 0 & 0 & K_{3}
\end{array}\right].
\end{equation}
The determinant of this matrix
\begin{equation}
\det\mathrm{H}=K_{1}K_{2}K_{3}\det\left[\begin{array}{ccc}
0 & 2K_{2} & -2K_{3}\\
-2K_{1} & 0 & 2K_{3}\\
2K_{1} & -2K_{2} & 0
\end{array}\right]=0\label{eq:p23}
\end{equation}
and the transformation of parameters in Eq. (\ref{eq:p9}) is not
invertible, since the rows and columns of the matrix in Eq. (\ref{eq:p23})
are linearly dependent. Thus $K_{1},K_{2},K_{3}$ cannot be determined
from knowledge of $p,q,r$.

\subsection{Estimating the quadratic invariants}

Owing to this, it remains to recover directly the invariants of the
TVG in Eq. (\ref{eq:p10}). Consider the more general problem of finding
quadratic invariants, given a vector field. We consider quadratic
invariants only, since any invariants must be closely tied to those
of the VG. As for the VG, these can only be identified up to affine
transformations, giving us families of invariants. We consider general
quadratic functions
\begin{multline}
C\left(x_{1},x_{2},x_{3}\right)=\frac{1}{2}d_{200}x_{1}^{2}+\frac{1}{2}d_{020}x_{2}^{2}+\frac{1}{2}d_{002}x_{3}^{2}+d_{110}x_{1}x_{2}+d_{011}x_{2}x_{3}+d_{101}x_{1}x_{3}+d_{100}x_{1}+d_{010}x_{2}+d_{001}x_{3}.\label{eq:p24}
\end{multline}
For $C$ constant in time
\begin{equation}
C'=\frac{\partial C}{\partial x_{1}}x_{1}'+\frac{\partial C}{\partial x_{2}}x_{2}'+\frac{\partial C}{\partial x_{3}}x_{3}'=0
\end{equation}
and, differentiating Eq. (\ref{eq:p24}) and substituting for the
vector field and collecting terms
\begin{multline}
C'=\left(d_{200}p+d_{020}q+d_{002}r\right)x_{1}x_{2}x_{3}+d_{110}px_{2}^{2}x_{3}+d_{101}px_{2}x_{3}^{2}+d_{110}qx_{3}x_{1}^{2}+d_{011}qx_{3}^{2}x_{1}+d_{011}rx_{1}x_{2}^{2}+d_{101}rx_{1}^{2}x_{2}\\
+\left(-d_{200}c+d_{020}c+d_{101}a-d_{011}b+d_{001}r\right)x_{1}x_{2}+\left(-d_{020}a+d_{002}a+d_{110}b-d_{101}c+d_{100}p\right)x_{2}x_{3}\\
+\left(-d_{002}b+d_{200}b+d_{011}c-d_{110}a+d_{010}q\right)x_{3}x_{1}+\left(-d_{101}b+d_{110}c\right)x_{1}^{2}+\left(-d_{110}c+d_{011}a\right)x_{2}^{2}+\left(-d_{011}a+d_{101}b\right)x_{3}^{2}\\
+\left(-d_{001}b+d_{010}c\right)x_{1}+\left(-d_{100}c+d_{001}a\right)x_{2}+\left(-d_{010}a+d_{100}b\right)x_{3}=0.\label{eq:p26}
\end{multline}
Since each of the terms are linearly independent in time, all coefficients
in the above equation must vanish. Considering first the coefficients
of $x_{2}^{2}x_{3}$ and $x_{3}x_{1}^{2}$, we have 
\begin{equation}
d_{110}p=d_{110}q=0.
\end{equation}
If both $p$ and $q$ were zero, then $r=0$, owing to the energy-conserving
constraint $p+q+r=0$. Such a model is linear, and excluding such
linear models from consideration leads to $d_{110}=0$. Similarly,
we have also $d_{101}=d_{011}=0$, and constants of motion of the
TVG do not possess mixed quadratic terms such as $x_{1}x_{2}$ etc.
Thus we are led to the following equations for parameters in Eq. (\ref{eq:p24}),
in matrix vector form
\begin{equation}
\left[\begin{array}{cccccc}
p & q & r & 0 & 0 & 0\\
-c & c & 0 & 0 & 0 & r\\
0 & -a & a & p & 0 & 0\\
b & 0 & -b & 0 & q & 0\\
0 & 0 & 0 & 0 & c & -b\\
0 & 0 & 0 & -c & 0 & a\\
0 & 0 & 0 & b & -a & 0
\end{array}\right]\left[\begin{array}{c}
d_{200}\\
d_{020}\\
d_{002}\\
d_{100}\\
d_{010}\\
d_{001}
\end{array}\right]=\left[\begin{array}{c}
0\\
0\\
0\\
0\\
0\\
0
\end{array}\right]\label{eq:p28}
\end{equation}
or, 
\begin{equation}
\mathrm{\mathrm{B}d=0}
\end{equation}
where $\mathrm{\mathrm{B}}\in\mathbb{R}^{7\times6}$ and $d=\left[\begin{array}{cccccc}
d_{200} & d_{020} & d_{002} & d_{100} & d_{010} & d_{001}\end{array}\right]^{T}$. That is, $\mathrm{d}\in\textrm{NULL}\left(\mathrm{B}\right)$, and
the number of independent quadratic invariants is governed by the
dimension of $\textrm{NULL}\left(\mathrm{B}\right)$. For two quadratic
invariants to exist, we must have $\dim\left(\textrm{NULL}\left(\mathrm{B}\right)\right)=2$.
From the rank-nullity theorem, $\dim\left(\textrm{NULL}\left(\mathrm{B}\right)\right)+\dim\left(\textrm{RANGE}\left(\mathrm{B}\right)\right)=6$,
so two quadratic invariants are inconsistent with more than $4$ linearly
independent rows (or columns) of the matrix $\mathrm{B}$.

\subsubsection*{General case}

We first treat the general case, with all of $p,q,r,a,b,c$ being
non-zero. From the last two equations
\begin{align}
d_{010} & =\frac{b}{a}d_{100}\nonumber \\
d_{001} & =\frac{c}{a}d_{100}\label{eq:p30}
\end{align}
and substituting these into the fourth equation
\begin{equation}
d_{002}=d_{200}+\frac{q}{a}d_{100}.
\end{equation}
The second equation yields
\begin{align}
d_{020} & =d_{200}-\frac{r}{c}d_{001}=d_{200}-\frac{r}{a}d_{100}\nonumber \\
 & =d_{200}+\frac{p+q}{a}d_{100}
\end{align}
where we have used Eq. (\ref{eq:p30}) and $p+q+r=0$. The first,
third, and fifth equations are linearly dependent with these that
have been used. Using these relations, theinvariants are of the form
\begin{align}
C\left(x_{1},x_{2},x_{3}\right) & =\frac{1}{2}d_{200}x_{1}^{2}+\frac{1}{2}d_{020}x_{2}^{2}+\frac{1}{2}d_{002}x_{3}^{2}+d_{100}x_{1}+d_{010}x_{2}+d_{001}x_{3}\nonumber \\
 & =\frac{1}{2}d_{200}x_{1}^{2}+\frac{1}{2}\left(d_{200}+\frac{p+q}{a}d_{100}\right)x_{2}^{2}+\frac{1}{2}\left(d_{200}+\frac{q}{a}d_{100}\right)x_{3}^{2}+d_{100}x_{1}+\frac{b}{a}d_{100}x_{2}+\frac{c}{a}d_{100}x_{3}
\end{align}
and rewritten in terms of $d_{200}$ and $d_{100}$
\begin{equation}
C\left(x_{1},x_{2},x_{3}\right)=\frac{1}{2}d_{200}\left(x_{1}^{2}+x_{2}^{2}+x_{3}^{2}\right)+d_{100}\left(\frac{p+q}{2a}x_{2}^{2}+\frac{q}{2a}x_{3}^{2}+x_{1}+\frac{b}{a}x_{2}+\frac{c}{a}x_{3}\right).
\end{equation}
Thus, the two invariants are
\begin{equation}
C_{1}\left(x_{1},x_{2},x_{3}\right)=\frac{1}{2}\left(x_{1}^{2}+x_{2}^{2}+x_{3}^{2}\right),\label{eq:p35}
\end{equation}
which is kinetic energy, and
\begin{align}
C_{2}\left(x_{1},x_{2},x_{3}\right) & =\frac{p+q}{2a}x_{2}^{2}+\frac{q}{2a}x_{3}^{2}+x_{1}+\frac{b}{a}x_{2}+\frac{c}{a}x_{3},\label{eq:p36}
\end{align}
which is related to the angular momentum squared as shown below. Substituting
$K_{i}y_{i}=x_{i}$ and Eqs. (\ref{eq:p9}) above
\begin{multline}
C_{2}=\frac{K_{2}^{2}-K_{1}^{2}}{2K_{1}h_{1}}K_{2}^{2}y_{2}^{2}+\frac{K_{3}^{2}-K_{1}^{2}}{2K_{1}h_{1}}K_{3}^{2}y_{3}^{2}+K_{1}y_{1}+\frac{K_{2}h_{2}}{K_{1}h_{1}}K_{2}y_{2}+\frac{K_{3}h_{3}}{K_{1}h_{1}}K_{3}y_{3}
\end{multline}
which can be written as 
\begin{align*}
K_{1}h_{1}C_{2} & =\frac{1}{2}\left(K_{1}^{2}y_{1}+h_{1}\right)^{2}+\frac{1}{2}\left(K_{2}^{2}y_{2}+h_{2}\right)^{2}+\frac{1}{2}\left(K_{3}^{2}y_{3}+h_{3}\right)^{2}-K_{1}^{2}\left(\frac{K_{1}^{2}y_{1}^{2}+K_{2}^{2}y_{2}^{2}+K_{3}^{2}y_{3}^{2}}{2}\right)-\left(h_{1}^{2}+h_{2}^{2}+h_{3}^{2}\right)
\end{align*}
and therefore
\begin{equation}
K_{1}h_{1}C_{2}=M-K_{1}^{2}E-\left(h_{1}^{2}+h_{2}^{2}+h_{3}^{2}\right).
\end{equation}
Since $\dot{C}_{2}=\dot{M}-K_{1}^{2}\dot{E}=0$, we have both $\dot{M}=\dot{E}=0$,
that is angular momentum and kinetic energy are conserved in the TVG,
as in the VG.

\subsubsection*{Subclasses of the TVG}

Further subclasses of the TVG, wherein one or more of its parameters
$p,q,r,a,b,c$ are zero, have been distinguished by GT (1999). From
the constraint that $p+q+r=0$, at least two of the quadratic coefficients
must be nonzero for the TVG to indeed be nonlinear. Thus, owing to
the symmetries in the model, the authors have focused on the situation
with $r=0$, since the cases with either $p=0$ or $q=0$ are analogous.
This yields nine distinct subclasses, whose constants of motion are
summarized in this section (see Table 1). Further details of the calculation
are provided in the Supplementary Information (SI). This model has
two free parameters in Eq. (\ref{eq:p24}) (SI Table 1) and thus two
independent quadratic invariants (Table 1), for each subclass. Furthermore,
in every subclass of this model, $\frac{1}{2}\left(x_{1}^{2}+x_{2}^{2}+x_{3}^{2}\right)$
is constant. This corresponds to energy conservation in the TVG. Analysis
of the last two degenerate cases where $x_{3}'=0$ is found in the
SI.

The existence of these two quadratic invariants is guaranteed by properties
of the matrix $\mathrm{B}$ in Eq. (\ref{eq:p30}). For example, it
can be shown that three rows (e.g. the first, third, and fifth) are
linearly dependent with the others. For the non-degenerate cases the
rank of this matrix is the number of independent rows, i.e. $4$.
The rank and the dimension of its null-space must sum to $6$, the
number of independent parameters we seek to estimate for the invariants.
Thus the null-space is two-dimensional, and there are two constants
of motion in general for this model. A similar conclusion is found
for the degenerate cases (SI). These constants of motion correspond
to surfaces denoted as $\mathscr{C}_{1}$ and $\mathscr{C}_{2}$,
which are plotted in Figures 1 and 2, for the subclasses and full
model respectively. In summary, we have found two quadratic invariants
for all variants of the TVG, rooted in the physics of the VG itself.

\pagebreak{}

Table 1: Expressions for constants of motion of the TVG in Eq. (\ref{eq:p10}).
In each subclass, $\frac{1}{2}\left(x_{1}^{2}+x_{2}^{2}+x_{3}^{2}\right)$
is conserved.

\begin{tabular}{|c|c|c|c|}
\hline 
No. &
Subclass &
$C_{1}$ &
$C_{2}$\tabularnewline
\hline 
\hline 
1 &
$r=0;b=c=0$ &
$\frac{1}{2}\left(x_{1}^{2}+x_{2}^{2}+x_{3}^{2}\right)$ &
$\frac{q}{2a}x_{3}^{2}+x_{1}$\tabularnewline
\hline 
2 &
$r=0;c=0$ &
$\frac{1}{2}\left(x_{1}^{2}+x_{2}^{2}+x_{3}^{2}\right)$ &
$\frac{q}{2a}x_{3}^{2}+x_{1}+\frac{b}{a}x_{2}$\tabularnewline
\hline 
3 &
$r=0;b=0$ &
$\frac{1}{2}\left(x_{1}^{2}+x_{2}^{2}+x_{3}^{2}\right)$ &
$\frac{q}{2a}x_{3}^{2}+x_{1}+\frac{c}{a}x_{3}$\tabularnewline
\hline 
4 &
$r=0$ &
$\frac{1}{2}\left(x_{1}^{2}+x_{2}^{2}+x_{3}^{2}\right)$ &
$\frac{q}{2a}x_{3}^{2}+x_{1}+\frac{b}{a}x_{2}+\frac{c}{a}x_{3}$\tabularnewline
\hline 
5 &
$a=b=c=0$ &
$\frac{1}{2}\left(x_{1}^{2}+\frac{p}{p+q}x_{3}^{2}\right)$ &
$\frac{1}{2}\left(x_{2}^{2}+\frac{q}{p+q}x_{3}^{2}\right)$\tabularnewline
\hline 
6 &
$b=c=0$ &
$\frac{1}{2}\left(x_{1}^{2}+\frac{p}{p+q}x_{3}^{2}-\frac{2a}{p+q}x_{1}\right)$ &
$\frac{1}{2}\left(x_{2}^{2}+\frac{q}{p+q}x_{3}^{2}+\frac{2a}{p+q}x_{1}\right)$\tabularnewline
\hline 
7 &
$c=0$ &
$\frac{1}{2}\left(x_{1}^{2}+\frac{p}{p+q}x_{3}^{2}-\frac{2a}{p+q}x_{1}-\frac{2b}{p+q}x_{2}\right)$ &
$\frac{1}{2}\left(x_{2}^{2}+\frac{q}{p+q}x_{3}^{2}+\frac{2a}{p+q}x_{1}+\frac{2b}{p+q}x_{2}\right)$\tabularnewline
\hline 
8 &
$r=0;a=b=c=0$ &
$\frac{1}{2}\left(x_{1}^{2}+x_{2}^{2}\right)$ &
$\frac{1}{2}x_{3}^{2}$\tabularnewline
\hline 
9 &
$r=0;a=b=0$ &
$\frac{1}{2}\left(x_{1}^{2}+x_{2}^{2}\right)$ &
$\frac{1}{2}x_{3}^{2}$\tabularnewline
\hline 
\end{tabular}

\pagebreak{}

The intersection of the two-dimensional surfaces $\mathscr{C}_{1}$,
$\mathscr{C}_{2}$ is a one dimensional manifold, giving oscillatory
periodic solutions if $q,p$ have opposite sign. We have solved for
the implicit functions describing these trajectories, in Table 2.
The numerically integrated solutions are plotted as solid black curves
in SI Figures 1 and 2, where we have taken $p,q$ to have opposite
sign. SI Fig 1 shows the periodic nature of the solutions with time,
for the different subclasses of the model and corresponding constants
of motion are plotted versus time in SI Fig 2, confirming their nature.
SI Table 2 lists the fixed points of these equations and their stability.
Each of the fixed points has a zero eigenvalue so, by the theorem
of Hartman and Grobman \cite{Guckenheimer1983}, the nonlinear dynamics
cannot be inferred from a linearized stability analysis.

\LyXZeroWidthSpace{}

Table 2: Intersections of the two constants of motion of the TVG,
yielding oscillatory solution curves since $p,q$ are of opposite
sign.

\begin{tabular}{|c|c|c|}
\hline 
No. &
Subclass &
$\mathscr{C}_{1}\cap\mathscr{C}_{2}$\tabularnewline
\hline 
\hline 
1 &
$r=0;b=c=0$ &
$\frac{1}{2}\left\{ \left(x_{1}-\frac{a}{q}\right)^{2}+x_{2}^{2}\right\} =C_{1}-\frac{a}{q}C_{2}+\frac{1}{2}\frac{a^{2}}{q^{2}}$\tabularnewline
\hline 
2 &
$r=0;c=0$ &
$\frac{1}{2}\left\{ \left(x_{1}-\frac{a}{q}\right)^{2}+\left(x_{2}-\frac{b}{q}\right)^{2}\right\} =C_{1}-\frac{a}{q}C_{2}+\frac{1}{2}\frac{a^{2}+b^{2}}{q^{2}}$\tabularnewline
\hline 
3 &
$r=0;b=0$ &
$\frac{1}{2}\left\{ \left(\frac{q}{2a}x_{3}^{2}+\frac{c}{a}x_{3}-C_{2}\right)^{2}+x_{2}^{2}+x_{3}^{2}\right\} =C_{1}$\tabularnewline
\hline 
4 &
$r=0$ &
$\frac{1}{2}\left\{ \left(\frac{q}{2a}x_{3}^{2}+\frac{b}{a}x_{2}+\frac{c}{a}x_{3}-C_{2}\right)^{2}+x_{2}^{2}+x_{3}^{2}\right\} =C_{1}$\tabularnewline
\hline 
5 &
$a=b=c=0$ &
$\frac{1}{2}\left(x_{2}^{2}-\frac{q}{p}x_{1}^{2}\right)=-\frac{q}{p}C_{1}+C_{2}$\tabularnewline
\hline 
6 &
$b=c=0$ &
$\frac{1}{2}\left(\frac{p-q}{2p}x_{1}^{2}+x_{2}^{2}+\frac{a}{p}x_{1}\right)=\frac{p-q}{2p}C_{1}+C_{2}$\tabularnewline
\hline 
7 &
$c=0$ &
$\frac{1}{2}\left(\frac{p-q}{2p}x_{1}^{2}+x_{2}^{2}+\frac{a}{p}x_{1}+\frac{b}{p}x_{2}\right)=\frac{p-q}{2p}C_{1}+C_{2}$\tabularnewline
\hline 
8 &
$r=0;a=b=c=0$ &
$\frac{1}{2}\left(x_{1}^{2}+x_{2}^{2}\right)=C_{1}$\tabularnewline
\hline 
9 &
$r=0;a=b=0$ &
$\frac{1}{2}\left(x_{1}^{2}+x_{2}^{2}\right)=C_{1}$\tabularnewline
\hline 
\end{tabular}

\subsection{Role of energy conservation}

The TVG possesses two constants of motion, corresponding to kinetic
energy and squared angular momentum. Any linear combination of these
quantities, together with an affine transformation, is conserved,
i.e.
\begin{equation}
\alpha_{1}E+\alpha_{2}M+\alpha_{3}
\end{equation}
is constant, for any $\alpha_{1},\alpha_{2},\alpha_{3}\in\mathbb{R}$.
This is because the transformation of state variables from the VG
to the TVG is smooth. Although the constants of motion of the TVG
can be related to those of the VG, they are not naturally rendered
as such (Table 1), owing to the non-invertibility of parameters in
going from VG to TVG. An important constraint in the above discussion
is $p+q+r=0$, arising from the inextricable link between these parameters
and the moments of inertia, as Eq. (\ref{eq:p9}) shows. If the TVG
is rooted in the physics of the VG, then it must conserve kinetic
energy. As described by \cite{Lakshmivarahan2008a}, energy conserving
LOMs must obey this constraint.

It is useful to examine models that have the form given in Eq. (\ref{eq:p10}),
yet do not conserve energy. Many LOMs possess this structure, even
where they do not satisfy energy conservation, for example Lorenz's
maximum simplification equations \cite{Lorenz1960,Lakshmivarahan2006}
resemble the Euler gyroscope (subclass 5 in Table 2) with the difference
being that in Lorenz's model $p+q+r\neq0$. Although these equations
do not conserve kinetic energy, they are known to enjoy two invariants
\cite{Lakshmivarahan2006}. In the case of this model none of the
invariants correspond to kinetic energy of the gyrostat, which is
obviously not conserved. Yet, it is clear from the equations that
the flow conserves volume, since the trace of the Jacobian of the
vector field of the TVG
\begin{equation}
\left[\begin{array}{ccc}
\frac{\partial x_{1}^{'}}{\partial x_{1}} & \frac{\partial x_{1}^{'}}{\partial x_{2}} & \frac{\partial x_{1}^{'}}{\partial x_{3}}\\
\frac{\partial x_{2}^{'}}{\partial x_{1}} & \frac{\partial x_{2}^{'}}{\partial x_{2}} & \frac{\partial x_{2}^{'}}{\partial x_{3}}\\
\frac{\partial x_{3}^{'}}{\partial x_{1}} & \frac{\partial x_{3}^{'}}{\partial x_{2}} & \frac{\partial x_{3}^{'}}{\partial x_{3}}
\end{array}\right]=\left[\begin{array}{ccc}
0 & px_{3}-c & px_{2}+b\\
qx_{3}+c & 0 & qx_{1}-a\\
rx_{2}-b & rx_{1}+a & 0
\end{array}\right]
\end{equation}
is zero, regardless of whether energy is conserved. However volume
conservation does not assure the existence of constants of motion.
This is evaluated further for those variants of the TVG that do not
conserve kinetic energy for which, in analogy with the subclasses
defined by \cite{Gluhovsky1999}, we limit our present analysis to
models possessing two or more quadratic terms.

\subsubsection*{General case}

The most general TVG without energy conservation possesses no other
constants of motion, as shown below. From Eq. (\ref{eq:p26}), we
first consider the part arising from mixed quadratic terms in the
constants of motion
\begin{align}
d_{110}\left(px_{2}^{2}x_{3}+qx_{3}x_{1}^{2}\right)+d_{101}\left(px_{2}x_{3}^{2}+rx_{1}^{2}x_{2}\right)+d_{011}\left(qx_{3}^{2}x_{1}+rx_{1}x_{2}^{2}\right)=0
\end{align}
where the last equality holds because these terms are linearly independent
of the others. Since there are at least two quadratic terms, more
than one of $p,q,r$ is non-zero. Then, we cannot in general have
$px_{2}^{2}x_{3}+qx_{3}x_{1}^{2}=0$ and therefore $d_{110}=0$. Similarly,
$d_{101}=d_{011}=0$. Here too there cannot be any mixed quadratic
terms in the quadratic constants of motion.

Thus, we are left with the same system of equations as before for
the parameters 
\begin{align}
d_{200}p+d_{020}q+d_{002}r & =0\nonumber \\
-d_{200}c+d_{020}c+d_{001}r & =0\nonumber \\
-d_{020}a+d_{002}a+d_{100}p & =0\nonumber \\
d_{200}b-d_{002}b+d_{010}q & =0\nonumber \\
d_{010}c-d_{001}b & =0\nonumber \\
-d_{100}c+d_{001}a & =0\nonumber \\
d_{100}b-d_{010}a & =0\label{eq:p42}
\end{align}
with the difference that that $p+q+r\neq0$. From the last two equations,
we obtain as before
\begin{align}
d_{010} & =\frac{b}{a}d_{100}\nonumber \\
d_{001} & =\frac{c}{a}d_{100},\label{eq:p43}
\end{align}
while the fifth equation is linearly dependent as before. Substituting
into the fourth equation we obtain
\begin{equation}
d_{002}=d_{200}+\frac{q}{b}d_{010}=d_{200}+\frac{q}{a}d_{100}\label{eq:p44}
\end{equation}
and from the second equation
\begin{align}
d_{020} & =d_{200}-\frac{r}{c}d_{001}=d_{200}-\frac{r}{a}d_{100}.\label{eq:p45}
\end{align}
 The third equation yields 
\begin{align}
-\left(d_{200}-\frac{r}{a}d_{100}\right)a+\left(d_{200}+\frac{q}{a}d_{100}\right)a+d_{100}p & =0
\end{align}
or equivalently
\begin{equation}
\left(p+q+r\right)d_{100}=0.
\end{equation}
If $p+q+r\neq0$ then $d_{100}=0$ and, from Eq. (\ref{eq:p43}),
$d_{010}=d_{001}=0$. Furthermore, we also obtain, from Eqs. (\ref{eq:p44})-(\ref{eq:p45}),
$d_{002}=d_{020}=d_{200}.$ Then, from the first equation 
\begin{equation}
\left(p+q+r\right)d_{200}=0
\end{equation}
or, $d_{200}=0$, and therefore also $d_{020}=d_{002}=0$. Thus, in
the absence of energy conservation, there are no constants of motion
in the general case.

\subsubsection*{Maximum simplification equations}

The above discussion assumed that $a,b,c\neq0$. Let us repeat the
above analysis for Lorenz's maximum simplification equations where
$a=b=c=0$ and $p+q+r\neq0$ with moreover $p,q,r\neq0$. The equations
for the coefficients become 
\begin{align*}
d_{200}p+d_{020}q+d_{002}r & =0\\
d_{001}r & =0\\
-d_{100}p & =0\\
d_{010}q & =0
\end{align*}
yielding $d_{100}=d_{010}=d_{001}=0$ and 
\begin{equation}
d_{200}p+d_{020}q+d_{002}r=0
\end{equation}
for one constraint in $d_{200},d_{020},d_{002}$. Thus, there are
two constants of motion for this model possessing only quadratic terms.
The maximum simplification equations have \cite{Lakshmivarahan2006}
\begin{align}
p & =-\left(\frac{1}{k^{2}}-\frac{1}{l^{2}+k^{2}}\right)kl\nonumber \\
q & =\left(\frac{1}{l^{2}}-\frac{1}{l^{2}+k^{2}}\right)kl\nonumber \\
r & =-\frac{1}{2}\left(\frac{1}{l^{2}}-\frac{1}{k^{2}}\right)kl\label{eq:p50}
\end{align}
with 
\begin{equation}
p+q+r=\frac{1}{2}\left(\frac{1}{l^{2}}-\frac{1}{k^{2}}\right)kl\neq0
\end{equation}
since $l\neq k$. For this case two quadratic invariants have been
identified
\begin{equation}
E=\frac{1}{4}\left(\frac{A^{2}}{l^{2}}+\frac{F^{2}}{k^{2}}+\frac{2}{k^{2}+l^{2}}G^{2}\right)
\end{equation}
and
\begin{equation}
V=\frac{1}{2}\left(A^{2}+F^{2}+2G^{2}\right)
\end{equation}
where, $A,F,G$ are the state variables \cite{Lorenz1960}. Each of
these invariants satisfy $d_{100}=d_{010}=d_{001}=0$. In the first
case 
\begin{equation}
d_{200}=\frac{1}{2l^{2}};d_{020}=\frac{1}{2k^{2}};d_{002}=\frac{1}{k^{2}+l^{2}}
\end{equation}
so that 
\begin{multline}
d_{200}p+d_{020}q+d_{002}r=\left\{ -\frac{1}{2l^{2}}\left(\frac{1}{k^{2}}-\frac{1}{l^{2}+k^{2}}\right)+\frac{1}{2k^{2}}\left(\frac{1}{l^{2}}-\frac{1}{l^{2}+k^{2}}\right)-\frac{1}{2}\frac{1}{k^{2}+l^{2}}\left(\frac{1}{l^{2}}-\frac{1}{k^{2}}\right)\right\} kl=0.
\end{multline}
Similarly, in the second case
\begin{equation}
d_{200}=1;d_{020}=1;d_{002}=2
\end{equation}
so that
\begin{multline}
d_{200}p+d_{020}q+d_{002}r=\left\{ -\left(\frac{1}{k^{2}}-\frac{1}{l^{2}+k^{2}}\right)+\left(\frac{1}{l^{2}}-\frac{1}{l^{2}+k^{2}}\right)-\left(\frac{1}{l^{2}}-\frac{1}{k^{2}}\right)\right\} kl=0.
\end{multline}
Thus the maximum simplification equations, together with the more
general subclass of which they are a part, exhibit oscillatory solutions.

\subsubsection*{Subclasses of the TVG}

The previous two sections demonstrated that whenever there is energy
conservation in the TVG, these equations also enjoy a second constant
of motion. This leads to oscillatory solutions, for each of the nine
subclasses of the TVG as well as the general case where all parameters
are nonzero. In the absence of energy conservation, this is not the
case and the dynamics admit richer possibilities. To show this, we
list the quadratic invariants for the nine different subclasses in
Table 3 (details in SI). Recall that the general case possesses no
quadratic invariants. Without energy conservation, for the different
subclasses the number of such invariants ranges from zero to two.
The last two degenerate subclasses possess two independent constants
(SI).
\begin{itemize}
\item Two invariants are enjoyed by those subclasses having either $a=b=c=0$
or $b=c=0$, or $a=b=0$ for the degenerate cases (Table 3). These
subclasses are the ones containing at most a single linear coefficient.
For example $b=c=0$ simplifies the equations for the coefficients
\begin{equation}
\left[\begin{array}{cccccc}
p & q & r & 0 & 0 & 0\\
0 & 0 & 0 & 0 & 0 & r\\
0 & -a & a & p & 0 & 0\\
0 & 0 & 0 & 0 & q & 0\\
0 & 0 & 0 & 0 & 0 & a\\
0 & 0 & 0 & 0 & -a & 0
\end{array}\right]\left[\begin{array}{c}
d_{200}\\
d_{020}\\
d_{002}\\
d_{001}\\
d_{010}\\
d_{001}
\end{array}\right]=\left[\begin{array}{c}
0\\
0\\
0\\
0\\
0
\end{array}\right]\label{eq:p58}
\end{equation}
so $d_{010}=d_{001}=0$, which leaves two independent equations
\begin{align}
d_{200}p+d_{020}q+d_{002}r & =0\nonumber \\
-d_{020}a+d_{002}a+d_{100}p & =0
\end{align}
in four unknowns, making for two independent invariants. With each
of $a=b=c=0$ this situation remains basically the same, leading to
the subclass that corresponds to Lorenz's maximum simplification equations,
which, as noted earlier, has periodic solutions.
\item Table 3 shows that a single constant of motion is held by subclasses
of the model having two nonzero linear coefficients. For example with
$c=0$ the equations become
\begin{equation}
\left[\begin{array}{cccccc}
p & q & r & 0 & 0 & 0\\
0 & 0 & 0 & 0 & 0 & r\\
0 & -a & a & p & 0 & 0\\
b & 0 & -b & 0 & q & 0\\
0 & 0 & 0 & 0 & 0 & -b\\
0 & 0 & 0 & 0 & 0 & a\\
0 & 0 & 0 & b & -a & 0
\end{array}\right]\left[\begin{array}{c}
d_{200}\\
d_{020}\\
d_{002}\\
d_{100}\\
d_{010}\\
d_{001}
\end{array}\right]=\left[\begin{array}{c}
0\\
0\\
0\\
0\\
0\\
0
\end{array}\right]\label{eq:p60}
\end{equation}
yielding $d_{001}=0$ and leaving four linearly independent equations
\begin{align}
d_{200}p+d_{020}q+d_{002}r & =0\nonumber \\
-d_{020}a+d_{002}a+d_{100}p & =0\nonumber \\
d_{200}b-d_{002}b+d_{010}q & =0\nonumber \\
d_{100}b-d_{010}a & =0\label{eq:p61}
\end{align}
in the remaining five parameters, and making for one constant of motion.
\item The skew-symmetric structure of the coefficient matrix
\begin{equation}
\left[\begin{array}{ccc}
0 & c & -b\\
-c & 0 & a\\
b & -a & 0
\end{array}\right]
\end{equation}

in the last three equations of Eq. (\ref{eq:p42}) confers a zero
determinant, so only two of these three equations are linearly independent.
In subclasses where $a,b,c\neq0$ the remainder of the equations are
linearly independent, making for six equations in six unknowns and
only a trivial solution to the invariants.
\end{itemize}
The invariants for the subclasses are plotted in Figure 3, along with
numerically integrated trajectories starting from the initial condition
$\mathrm{x}\left(0\right)=\left[\begin{array}{ccc}
1 & 1 & 1\end{array}\right]^{T}$. In general, only those subclasses enjoying two invariants have periodic
solutions from any initial condition. The corresponding time-series
of $C_{1}$ and $C_{2},$where they exist, are shown in SI Fig 3.
Figure 3 of the paper illustrates three subclasses (2,3,7) where dynamics
resides on a two-dimensional manifold, as the result of enjoying a
single constant of motion. The absence of periodicity of these solutions
is apparent from the relative thickness of these trajectories in the
figure. Since the dynamics of these subclasses is circumscribed by
the Poincar$\acute{\textrm{e}}$-Bendixson theorem \cite{Guckenheimer1983},
we will not consider these further. Figure 3 also shows a subclass
(4) that, like the general case, has no invariants as a result of
all the linear coefficients being nonzero. Thus, relaxing the energy
conservation constraint opens the door to much richer dynamics of
the TVG, as illustrated below.

\pagebreak{}

Table 3: Constants of motion for $p+q+r\neq0$. Subclasses with three
linear coefficients have no invariants, those with two linear coefficients
have one quadratic invariant, and those with one or no linear coefficient
have two independent quadratic invariants. The number of independent
quadratic invariants is given by $k$.

\begin{tabular}{|c|c|c|c|c|c|}
\hline 
No. &
Subclass &
$C_{1}$ &
$C_{2}$ &
$C_{3}$ &
$k$\tabularnewline
\hline 
\hline 
1 &
$r=0;b=c=0$ &
$\frac{1}{2}\left(x_{1}^{2}-\frac{p}{q}x_{2}^{2}-\frac{p}{q}x_{3}^{2}\right)$ &
$-\frac{p}{2a}x_{3}^{2}+x_{1}$ &
 &
$2$\tabularnewline
\hline 
2 &
$r=0;c=0$ &
$\frac{1}{2}\left(x_{1}^{2}-\frac{p}{q}x_{2}^{2}-2\frac{a}{q}x_{1}-2\frac{b}{q}x_{2}\right)$ &
 &
 &
$1$\tabularnewline
\hline 
3 &
$r=0;b=0$ &
$-\frac{1}{2}\frac{p}{a}x_{3}^{2}+x_{1}+\frac{c}{a}x_{3}$ &
 &
 &
$1$\tabularnewline
\hline 
4 &
$r=0$ &
 &
 &
 &
$0$\tabularnewline
\hline 
5 &
$a=b=c=0$ &
$\frac{1}{2}\left(x_{1}^{2}-\frac{p}{r}x_{3}^{2}\right)$ &
$\frac{1}{2}\left(x_{2}^{2}-\frac{q}{r}x_{3}^{2}\right)$ &
 &
$2$\tabularnewline
\hline 
6 &
$b=c=0$ &
$\frac{1}{2}\left(x_{1}^{2}-\frac{p}{q+r}x_{2}^{2}-\frac{p}{q+r}x_{3}^{2}\right)$ &
$\frac{1}{2}\left(\frac{pr}{\left(q+r\right)a}x_{2}^{2}-\frac{pq}{\left(q+r\right)a}x_{3}^{2}+2x_{1}\right)$ &
 &
$2$\tabularnewline
\hline 
7 &
$c=0$ &
$\frac{1}{2}\left(x_{1}^{2}-\frac{p}{q}x_{2}^{2}-2\frac{a}{q}x_{1}-2\frac{b}{q}x_{2}\right)$ &
 &
 &
$1$\tabularnewline
\hline 
8 &
$r=0;a=b=c=0$ &
$\frac{1}{2}\left(x_{1}^{2}-\frac{p}{q}x_{2}^{2}\right)$ &
$\frac{1}{2}x_{3}^{2}$ &
 &
$2$\tabularnewline
\hline 
9 &
$r=0;a=b=0$ &
$\frac{1}{2}\left(x_{1}^{2}-\frac{px_{30}-c}{qx_{30}+c}x_{2}^{2}\right)$ &
$\frac{1}{2}x_{3}^{2}$ &
 &
$2$\tabularnewline
\hline 
\end{tabular}

\pagebreak{}

\section{Chaos in the TVG}

Where quadratic invariants do not exist for these 3-dimesional flows
(Subclasses 4 and the general case) the dynamics are rich, and include
irregular dynamics as well as chaos. We examine this further for subclass
4, for which we have examined a large ensemble of varying parameters
and initial conditions using Latin hypercube sampling. Examples of
irregular trajectories, their orbits and time-series of $x_{1}$,
are illustrated in SI Figs 4-5. We plot some of these cases in Figures
4-6. A minimal chaotic model in the TVG without energy conservation
has
\begin{align}
x_{1}' & =px_{2}x_{3}+bx_{3}-cx_{2}\nonumber \\
x_{2}' & =\left(-p+\delta\right)x_{3}x_{1}+cx_{1}-ax_{3}\nonumber \\
x_{3}' & =ax_{2}-bx_{1}\label{eq:p63}
\end{align}
where $q=-p+\delta$, so that $E'=\delta x_{1}x_{2}x_{3}$. Recall
that with $\delta\neq0$, there is no energy conservation. Since there
are three linear feedbacks, this has no quadratic invariants (Table
3). Figure 4 shows the changing orbit sequence, for changing $\delta$,
including $\delta=0$, for fixed $a$, $b$, $c$, $p$, and $q$
calculated as $q=-p+\delta$. Corresponding time-series of $x_{1}$
and $E$ are shown in Figures 5-6. The dynamics can become chaotic
for nonzero $\delta$. Energy is conserved only for $\delta=0$ (Figure
6).

The particular example of $a=-0.67$, $b=0.18$, $c=0.70$, $p=0.76$,
and $q=-p+\delta$, is taken up further in Figure 7, which shows transient
and stationary orbits for uniform increase in $\delta$, from $-0.25$
to $0$. Poincar$\acute{\textrm{e}}$ sections (Figure 8) confirm
the appearance of chaos for each of these cases except $\delta=-0.063$
and $\delta=0$, and this is also confirmed by corresponding power
spectral densities having continuous peaks (Figure 9), as well as
positive finite-time Lyapunov exponents (Figure 10). Corresponding
time-series are shown in SI Fig 6. These last examples serve to demonstrate
that chaos can appear in the volume conserving TVG.

By analogy with subclass 4, for the full TVG with three quadratic
terms we can write a model that admits chaos
\begin{align}
x_{1}' & =px_{2}x_{3}+bx_{3}-cx_{2}\nonumber \\
x_{2}' & =qx_{3}x_{1}+cx_{1}-ax_{3}\nonumber \\
x_{3}' & =\left(-p-q+\delta\right)x_{1}x_{2}+ax_{2}-bx_{1}\label{eq:p64}
\end{align}
that also has non-conserved $E'=\delta x_{1}x_{2}x_{3}$ and no invariants
in case of nonzero $\delta$ owing to three linear feedbacks. These
systems in Eqs. (\ref{eq:p63})-(\ref{eq:p64}) both preserve volume
as the flow evolves.

In summary, a necessary condition for chaos in the volume-conserving
TVG is the presence of three linear feedbacks (subclass 4 and the
general case) along with two quadratic nonlinearities with nonzero
$\delta$ as in subclass 4. We have not considered models with a single
quadratic nonlinearity.

\pagebreak{}

\begin{landscape}

Table 4: Summary of the major symbols used.

\begin{tabular}{|c|c|}
\hline 
Symbol &
Definition\tabularnewline
\hline 
\hline 
$y_{1},y_{2},y_{3}$ &
angular velocity of carrier body\tabularnewline
\hline 
$K_{1}^{2},K_{2}^{2},K_{3}^{2}$ &
principal moments of inertia of the gyrostat\tabularnewline
\hline 
$h_{1},h_{2},h_{3}$ &
angular momentum of the rotor relative to the carrier\tabularnewline
\hline 
$E$ &
kinetic energy: $E=\frac{1}{2}\sum_{i=1}^{3}K_{i}^{2}y_{i}^{2}$\tabularnewline
\hline 
$M$ &
squared angular momentum: $M=\frac{1}{2}\sum_{i=1}^{3}\left(K_{i}^{2}y_{i}+h_{i}\right)^{2}$\tabularnewline
\hline 
$x_{1},x_{2},x_{3}$ &
state variables of the TVG: $x_{1}=K_{1}y_{1},x_{2}=K_{2}y_{2},x_{3}=K_{3}y_{3}$\tabularnewline
\hline 
$p,q,r$ &
quadratic coefficients of TVG dynamics: $p=K_{2}^{2}-K_{3}^{2},q=K_{3}^{2}-K_{1}^{2},r=K_{1}^{2}-K_{2}^{2}$\tabularnewline
\hline 
$a,b,c$ &
linear coefficients of TVG dynamics: $a=K_{1}h_{1},b=K_{2}h_{2},c=K_{3}h_{3}$\tabularnewline
\hline 
$C_{i}\left(y_{1},y_{2},\ldots y_{n}\right)$ &
invariants of the system $\mathrm{\dot{y}}=\mathrm{g}\left(\mathrm{y}\right)$
, with $\mathrm{y}=\left[\begin{array}{cccc}
y_{1} & y_{2} & \ldots & y_{n}\end{array}\right]^{T}\in\mathbb{R}^{n}$ and $\mathrm{g}:\mathbb{R}^{n}\rightarrow\mathbb{R}^{n}$\tabularnewline
\hline 
$\mathrm{\mathrm{H}}$ &
Jacobian of parameter transformation involved in VG->TVG\tabularnewline
\hline 
$d_{200},d_{020},d_{002},d_{110},d_{011},d_{101},d_{100},d_{010},d_{001}$ &
coefficients of quadratic invariants of TVG\tabularnewline
\hline 
$\mathrm{d}$ &
vector of potentially non-zero coefficients: $d=\left[\begin{array}{cccccc}
d_{200} & d_{020} & d_{002} & d_{100} & d_{010} & d_{001}\end{array}\right]^{T}$\tabularnewline
\hline 
$\mathrm{B}$ &
matrix defining linear homogeneous equations satisfied by coefficients:
$\mathrm{Bd=0}$\tabularnewline
\hline 
$C_{1}\left(x_{1},x_{2},x_{3}\right)$ &
first quadratic invariant of TVG (where it exists)\tabularnewline
\hline 
$C_{2}\left(x_{1},x_{2},x_{3}\right)$ &
second quadratic invariant of TVG (where it exists)\tabularnewline
\hline 
$\mathscr{C}_{1}$ and $\mathscr{C}_{2}$ &
surfaces describing the respective quadratic invariants\tabularnewline
\hline 
$\delta$ &
$p+q+r$, which is nonzero in the absence of energy conservation\tabularnewline
\hline 
\end{tabular}

\end{landscape}

\pagebreak{}

\section{Summary and Discussion}

The Volterra gyrostat (VG) and its transformation (TVG) appear ubiquitously
within the structures of many low-lower models (LOMs) \cite{Gluhovsky1997,Gluhovsky2002},
and this paper provides a characterization of their quadratic invariants
for each of the special cases discussed in \cite{Gluhovsky1999} (Section
2). The present paper provides an explicit account of the expressions
for quadratic invariants in the TVG and, given the close link between
the number of invariants and the structure of these models, such inquiries
are also relevant to identifying the conditions of chaotic dynamics
in such models. We have shown how the number of invariants depends
on the presence of the energy conservation constraint, as well as
the number of linear feedback terms. The study shows that, since the
TVG arises from a smooth transformation of variables it possesses
two quadratic invariants in the presence of energy conservation, despite
the parameter transformation not being invertible. Energy conservation
corresponds to the quadratic coefficients in the TVG summing to zero.
These invariants correspond to the same conservation laws as the VG
even though they are not explicitly rendered as such; wherein energy
and angular momentum and any affine transformation are conserved,
giving periodic dynamics from any initial condition.

It is also shown that energy conservation is central to existence
of two invariants of the TVG, and relaxing this constraint admits
wider possibilities, from zero to two invariants. Such models without
energy conservation also appear as the fundamental core of LOMs in
weather and climate (e.g., \cite{Lorenz1960}), and their characterization
becomes important. The number of invariants in these more general
models is related to the number of linear terms, with subclasses having
three linear feedback terms possessing none. It is shown, in the absence
of the energy conservation constraint, that subclasses with three
linear coefficients have no invariants, those with two linear coefficients
have one quadratic invariant, and those with one or no linear coefficient
have two independent quadratic invariants (Section 2). Chaos can arise
in the volume conserving flows with three linear feedback terms (Section
3).

The core of the paper (Section 2) is focused on estimating quadratic
invariants in three state variables. We estimate the number of independent
invariants that are consistent with the evolution equations of various
subclasses of the TVG. Our study focuses on the gyrostat with two
or more nonlinear terms. This reduces to estimating $6$ coefficients,
once we observe that for the TVG with atleast two nonlinear terms
the quadratic invariants cannot posses any mixed quadratic term. Then
the number of quadratic invariants is related to the number of linearly
independent solutions to a system of homogeneous linear equations
(in these $6$ unknown coefficients). These linear equations come
from the consistency of the quadratic invariants with the evolution
equations of the TVG. The analysis is repeated with and without the
energy conservation constraint being present, clearly showing the
relevance of energy conservation to the existence of a second invariant.

For these three-dimensional models the recognition of invariants,
where they exist, obviates the need for explicit integration of the
equations for understanding asymptotic as well as transient dynamics.
Numerical investigations are of course important to identify the possibility
of chaos, when no invariants exist. While we have not examined the
existence of higher invariants of the model, the consistency of the
numerical simulations with the number of quadratic invariants confirms
the approach. We have numerically studied particular examples giving
chaos when $\delta\equiv p+q+r\neq0$. Circumscribing the possibilities
for dynamics when this sum of quadratic coefficients $\delta$ is
nonzero is an open problem. Since chaos cannot arise when $\delta=0$,
it might serve naturally as a bifurcation parameter inviting further
study. It is also important to consider the origin of nonzero $\delta$
in the conservative core of models having these structures (e.g. \cite{Lorenz1960,Leipnik1981}).

Previous authors have considered simple chaotic flows in three dimensions
, where the simplicity of the model is usually characterized by fewer
number of distinct terms defining the vector field. For example, \cite{Sprott1994}
considered vector fields having five different terms (with two of
them being nonlinear) or six terms (with one of them being nonlinear).
These studies have illuminated the algebraically simplest chaotic
models, most of which are dissipative, but there are examples of volume
conserving flows too \cite{Sprott1994,Sprott2000,Heidel2007}. In
fact, the simplest volume conserving chaotic flow in three dimensions
has only four terms, of which two are nonlinear \cite{Heidel2007}.
In comparison subclass 4 in Eq. (\ref{eq:p63}) has eight terms, of
which two are nonlinear, and thus is in no respect simple. Given that
the present inquiry is confined to models having the gyrostat structure,
and that we have not considered those subclasses with a single nonlinearity,
we do not seek to identify simple chaotic models among more general
vector fields. The importance of the models in Eqs. (\ref{eq:p63})-(\ref{eq:p64})
lies in their origin in the gyrostat equations and thus ubiquity in
the structure of LOMs. Therefore, the simplest chaotic flows within
the family of gyrostats, and coupled gyrostats, merit inquiry despite
them not being cataloged among the simplest chaotic flows across more
general vector fields.

The famous LOM developed by Lorenz \cite{Lorenz1963} of the Rayleigh-Benard
problem of convective overturning in a fluid heated from below has
been interpreted by Gluhovsky and Tong \cite{Gluhovsky1999} as a
forced-dissipative version of subclass 1 of the TVG where $p+q+r=0$.
The authors \cite{Gluhovsky1999} show that this subclass has two
integrals of motion. In the present study, we have shown that two
integrals of motion are held by this subclass even if $p+q+r\neq0$.
The model of Lorenz has the same symmetry as its conservative core
described here \cite{Lorenz1963,Gilmore2007}, but its dynamics is
very different \cite{Gluhovsky1999}, and for the LOM of \cite{Lorenz1963}
the conditions of chaotic dynamics arise from the combined effects
of dissipation, which collapses volumes of initial conditions in phase
space, and effects of forcing.

Analogously, for each of the subclasses of the TVG, a rich collection
of quadratic LOMs can be conceived by extending these models to their
forced and dissipative counterparts, whose dynamics might depart significantly
from the models elucidated here. In general, with forcing and dissipation
we might expect chaotic behavior to arise for each of the subclasses,
regardless of the number of linear feedbacks. Moreover, subclass 4
and the most general case too will present chaos when forcing and
dissipation are included, but the nature of the orbits will surely
differ from those examined here. Since these LOMs shall in general
depart from the TVG in respect of forcing, dissipation, as well as
the presence of the energy conservation constraint, it is of interest
to study the role of each of these factors on the resulting possibilities
for dynamics in forced dissipative systems, and compare the mechanisms
giving rise to chaos in each case.

Furthermore, the concept of VG with linear feedback has been extended
to the generalized Volterra gyrostat having nonlinear feedback terms,
which also arise naturally in low-order models \cite{Lakshmivarahan2008a}.
It is of considerable interest to extend the present analysis to these
models, to study their various invariants and the conditions permitting
chaos. Ultimately, the TVG appears in low-order models not singly
but as part of a system of a variety of different subclasses of the
TVG that are coupled with each other. The underlying conservation
laws, when they are stripped of forcing and dissipation, can generally
possess quadratic invariants. Studying the invariants of systems of
coupled TVGs, involving both linear and nonlinear feedback terms,
is therefore a promising line of future inquiry. When derived from
conservation laws we can expect such low-order models to possess at
most two (or a few) quadratic invariants. Considering the more general
problem of invariants in systems of coupled gyrostats, as well as
the conditions for chaos in these models can help illuminate the behavior
of low-order models as well as the infinite-dimensional systems that
they describe.

\section*{Acknowledgments}

The authors are grateful to Frank Kwasniok, Vishal Vasan, and anonymous
reviewers for helpful suggestions.

\section*{Declarations of interest}

The authors have no competing interests to declare. 

\pagebreak{}

\bibliographystyle{ieeetr}
\bibliography{1D__Dropbox_0_ClimatePrediction_Paper3_Gyrostat_Paper1_conservative_VG_constants}

\pagebreak{}

\begin{figure}
\includegraphics[scale=0.8]{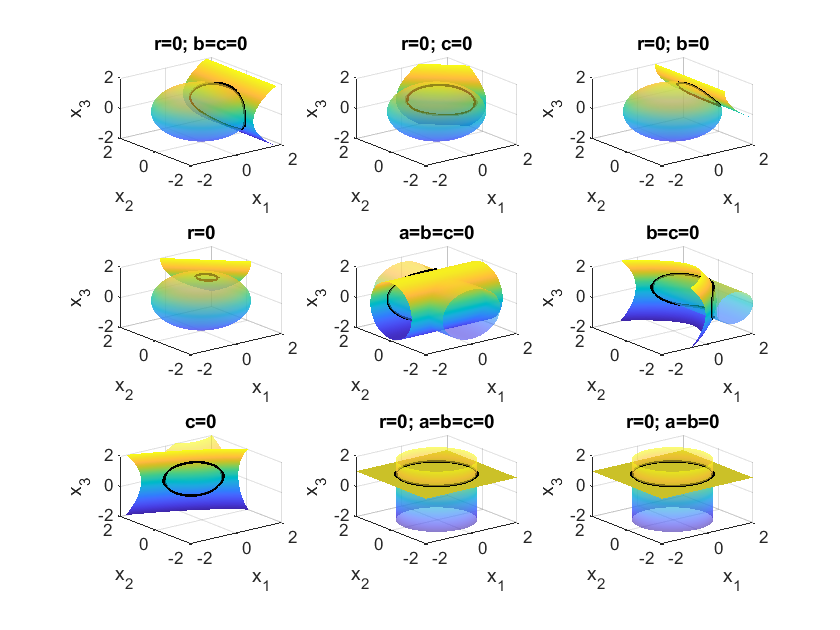}

\caption{The two quadratic invariants for the TVG, for the subclasses ordered
as in Table 1, and following \cite{Gluhovsky1999}. The solid black
curve shows corresponding solution trajectories of $\mathrm{x}$,
from numerical integration. Results are shown for the initial condition
$\mathrm{x}\left(0\right)=\left[\protect\begin{array}{ccc}
1 & 1 & 1\protect\end{array}\right]^{T}$. These plots assume energy conservation $p+q+r=0$, giving rise to
periodic solutions.}
\end{figure}

\begin{figure}
\includegraphics[scale=0.7]{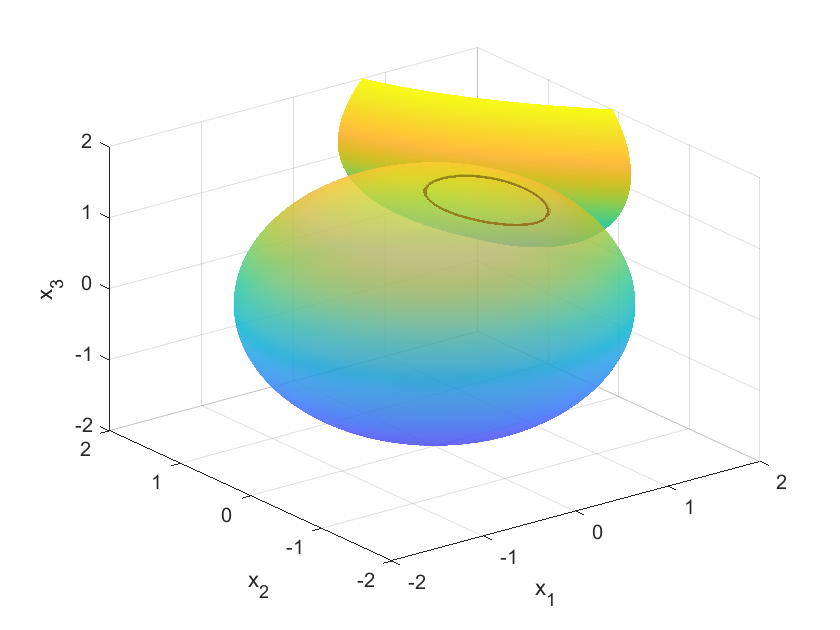}

\caption{The two quadratic invariants for the Volterra gyrostat, in Eqs. (\ref{eq:p35})
and (\ref{eq:p36}). The solid black curve shows the solution trajectories
of $\mathrm{x}$, from numerical integration, for this general case.
Results are shown for the initial condition $\mathrm{x}\left(0\right)=\left[\protect\begin{array}{ccc}
1 & 1 & 1\protect\end{array}\right]^{T}$.}
\end{figure}

\begin{figure}
\includegraphics[scale=0.56]{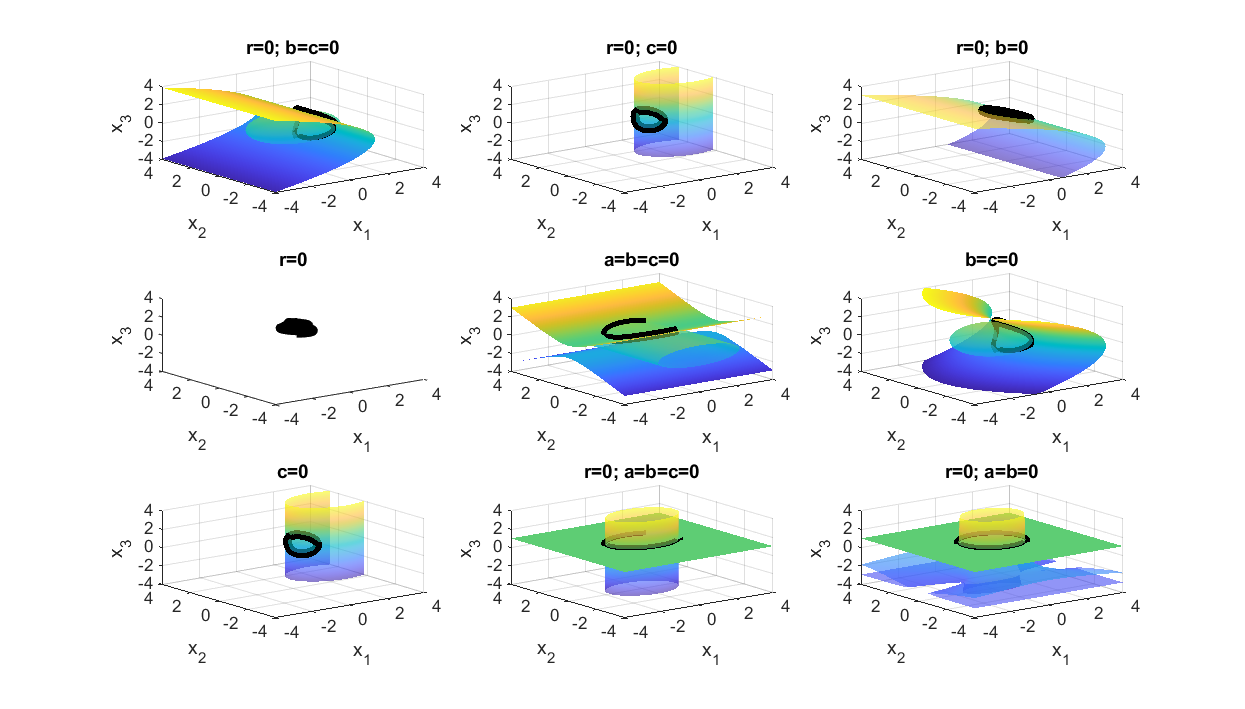}

\caption{Constants of motion for the TVG for the subclasses ordered as in Table
3 in the absence of energy conservation, i.e. $p+q+r\protect\neq0$.
The solid black curve shows corresponding solution trajectories of
$\mathrm{x}$, from numerical integration. Results are shown for the
initial condition $\mathrm{x}\left(0\right)=\left[\protect\begin{array}{ccc}
1 & 1 & 1\protect\end{array}\right]^{T}$. Only those subclasses with two quadratic invariants (1,5,6,8,9)
have periodic solutions.}
\end{figure}

\pagebreak{}

\begin{landscape}

\begin{figure}
\includegraphics[scale=0.5]{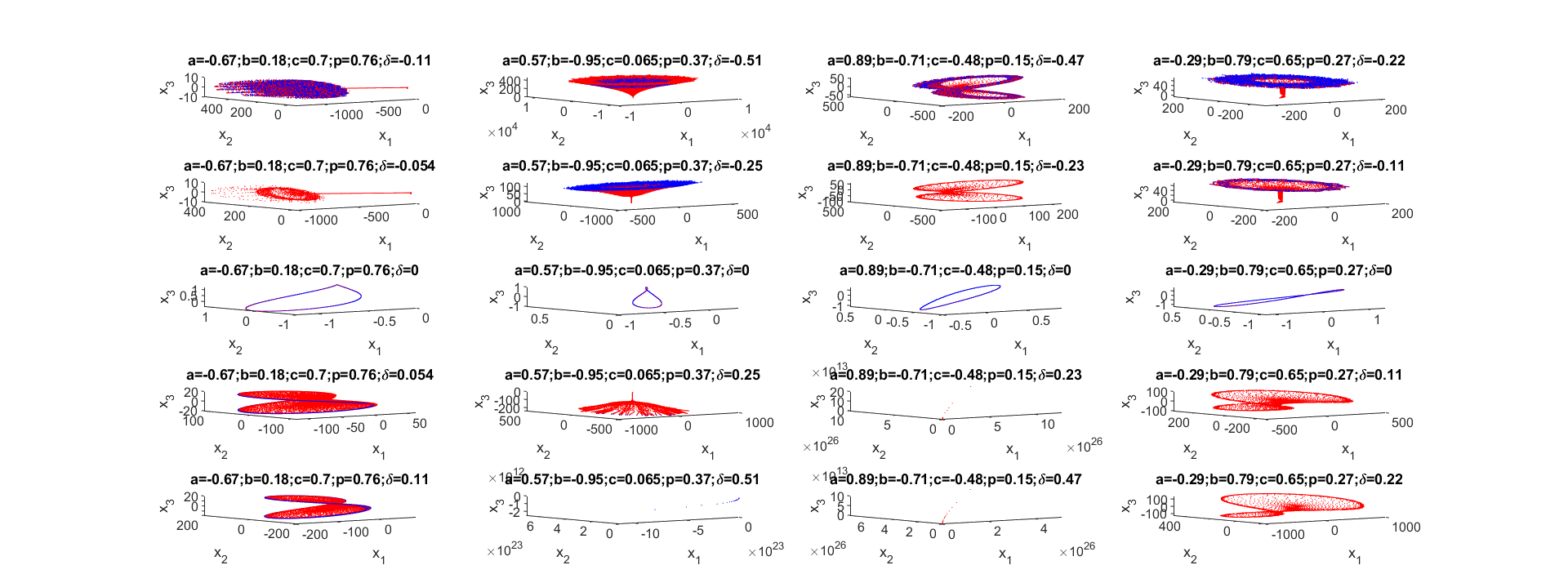}

\caption{Orbits of subclass 4 ($r=0$) without energy conservation, with each
column showing results for fixed $a$, $b$, $c$, $p$, and $q$
is calculated for each panel as $q=-p+\delta$, with $\delta$ varying
by rows. Transient dynamics is shown in red and stationary orbit in
blue. Corresponding time-series of $x_{1}$ and $E$ are shown in
Figures 5-6.}
\end{figure}

\pagebreak{}

\begin{figure}
\includegraphics[scale=0.5]{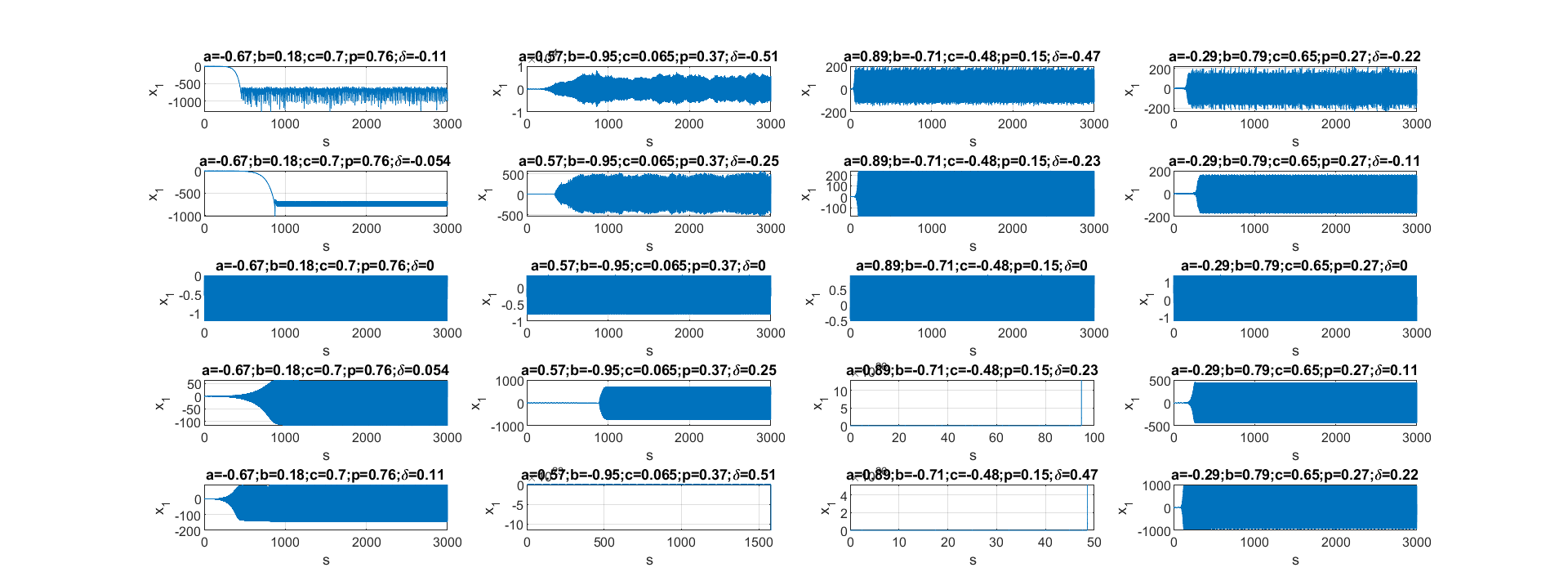}

\caption{Time-series of $x_{1}$ for the orbits in Figure 4.}
\end{figure}

\pagebreak{}

\begin{figure}
\includegraphics[scale=0.5]{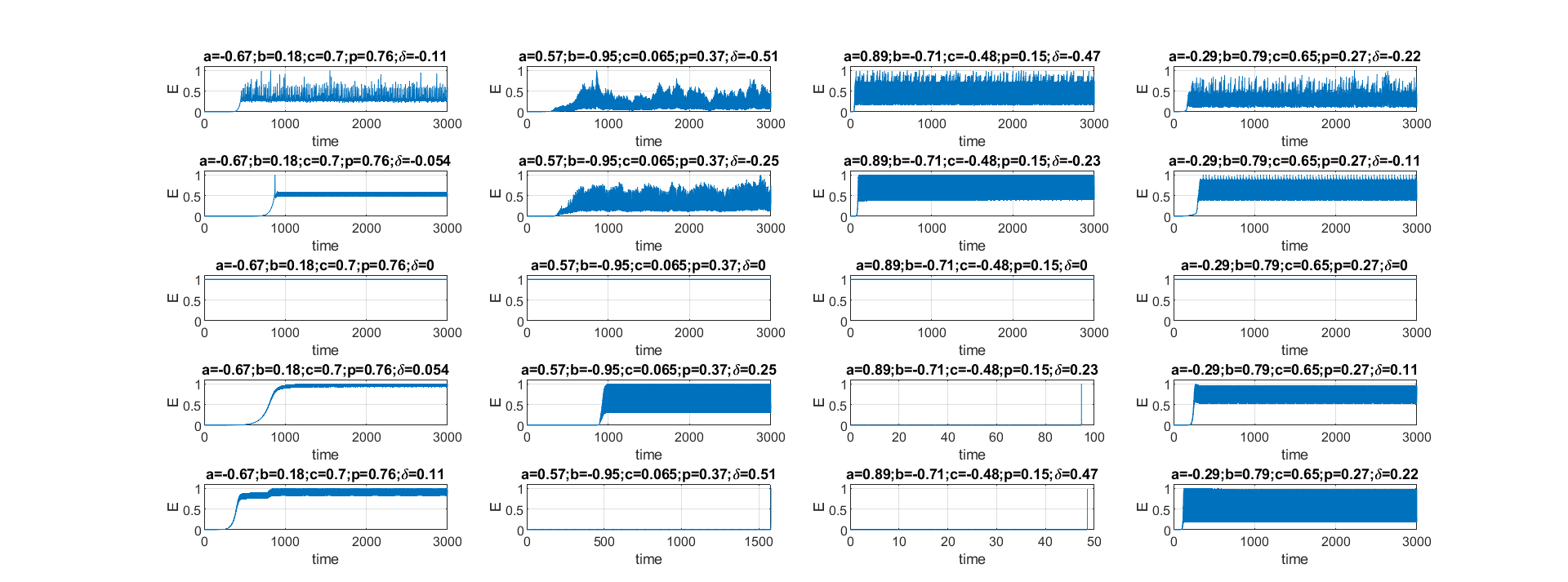}

\caption{Time-series of $E$ (normalized by its maximum value) for the orbits
in Figure 4. $E$ is constant only for $\delta=0$ in the middle row.}
\end{figure}

\pagebreak{}

\end{landscape}

\begin{figure}
\includegraphics[scale=0.4]{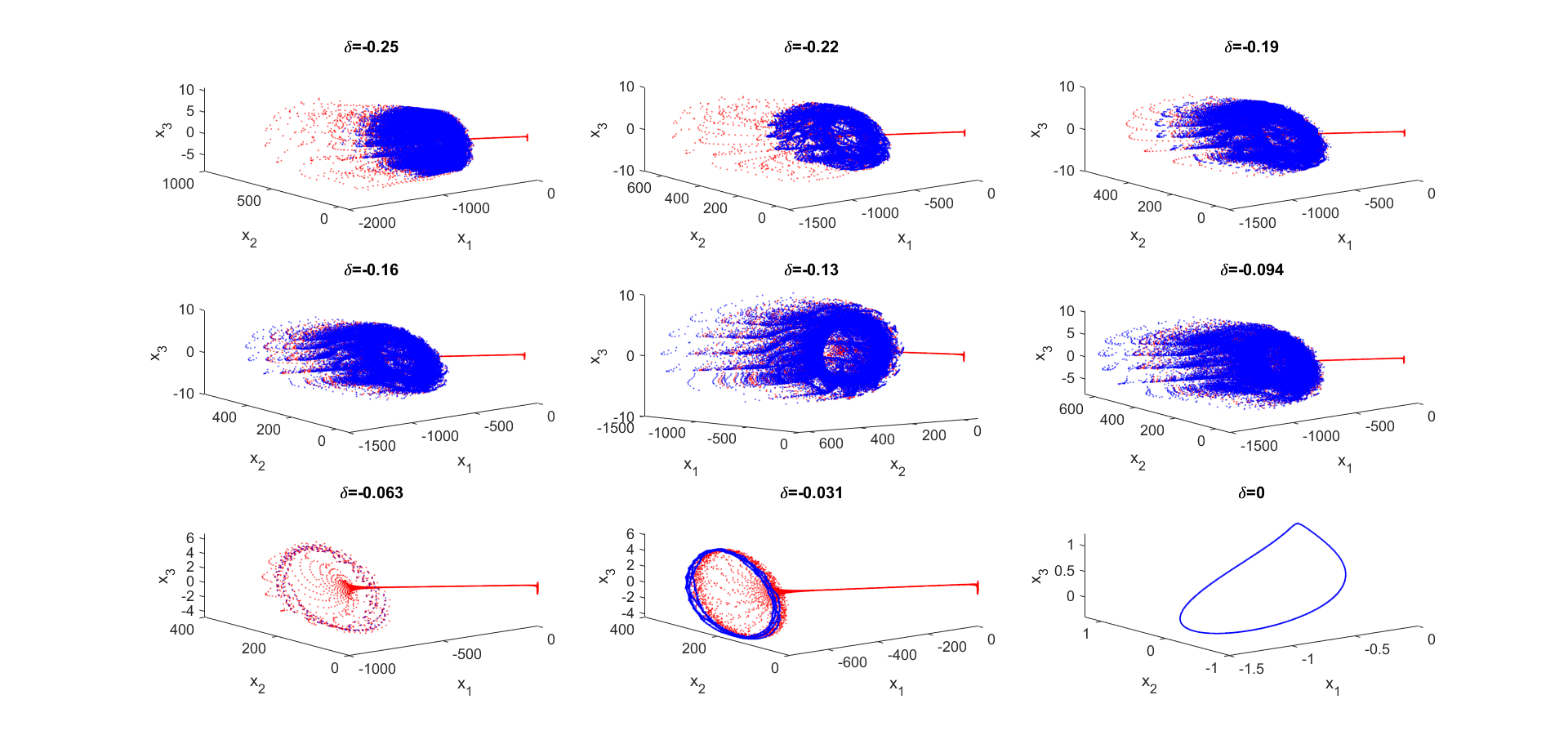}

\caption{Orbits of subclass 4 $\left(r=0\right)$ without energy conservation,
having $a=-0.67$, $b=0.18$, $c=0.70$, $p=0.76$, and $q=-p+\delta$,
as parameter $\delta$ is varied. Since $p+q=\delta$, energy conservation
occurs only for $\delta=0$. Successive panels show results for uniform
increments of $\delta$, where the stationary sequence is shown in
blue. Chaos occurs in each of these cases except $\delta=-0.063$
and $\delta=0$. Corresponding Poincar$\acute{\textrm{e}}$ sections
are in Figure 8, power spectra in Figure 9, and Lyapunov exponents
in Figure 10. SI Fig 6 shows time-series of $x_{1}$.}
\end{figure}

\pagebreak{}
\begin{figure}
\includegraphics[scale=0.4]{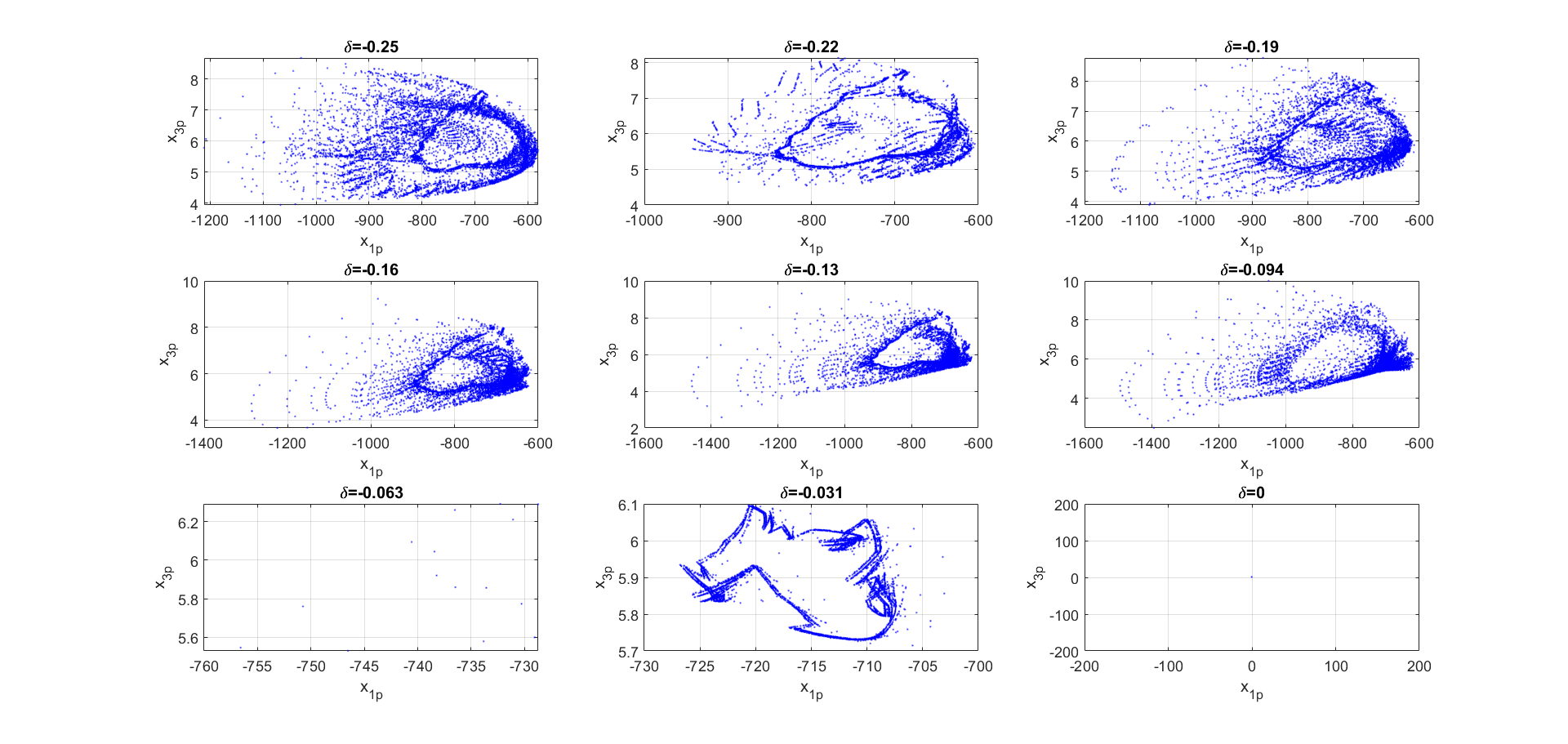}

\caption{Poincar$\acute{\textrm{e}}$ sections of the stationary sequences
(subclass 4) in Figure 7, demonstrating that chaos can occur for certain
parameter values.}
\end{figure}

\pagebreak{}

\begin{figure}
\includegraphics[scale=0.4]{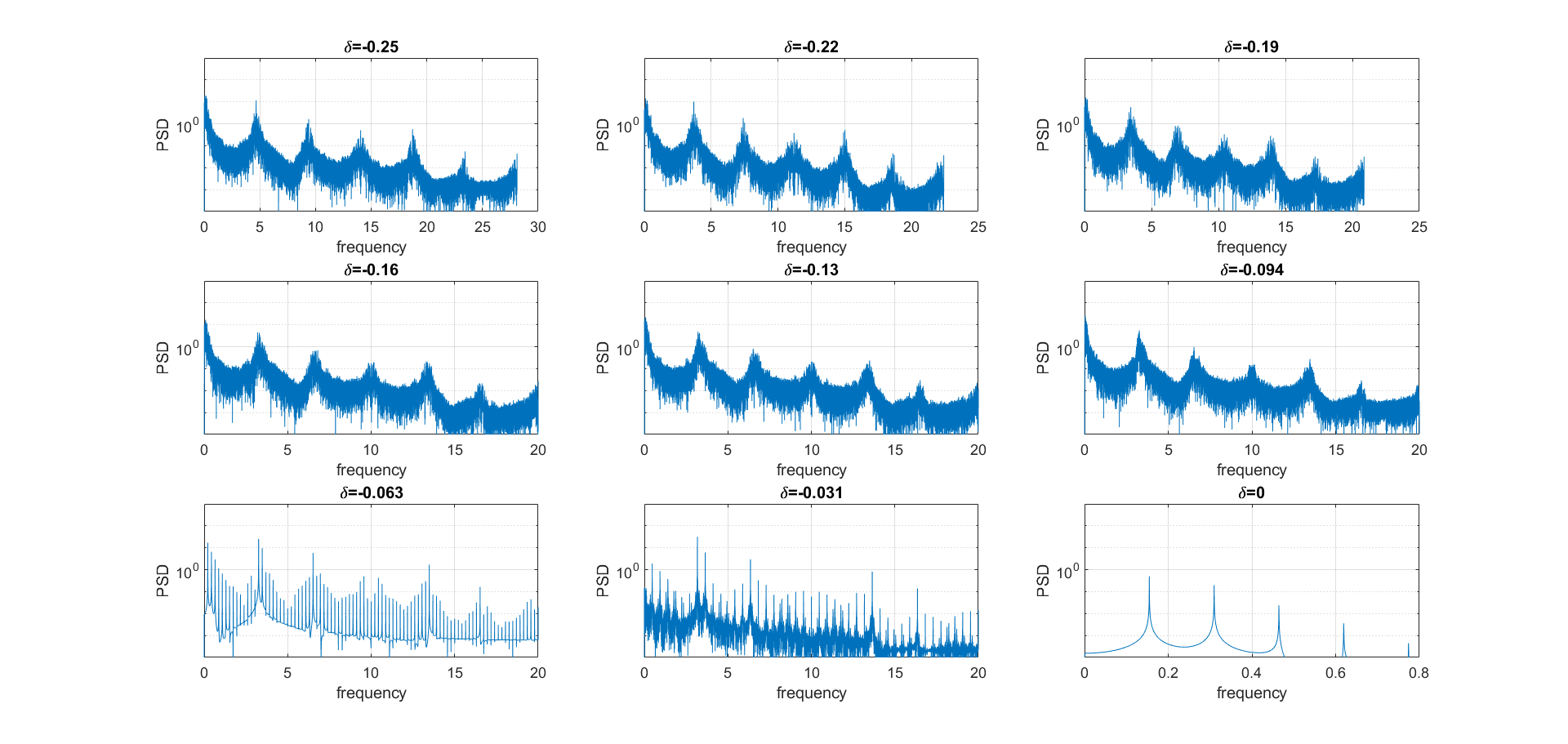}

\caption{Power spectral density (log-scale) versus frequency (linear-scale)
for stationary orbits in Figure 7, for subclass 4 without energy conservation,
showing broadband spectra associated with chaos for certain parameter
values.}
\end{figure}

\pagebreak{}

\begin{figure}
\includegraphics[scale=0.4]{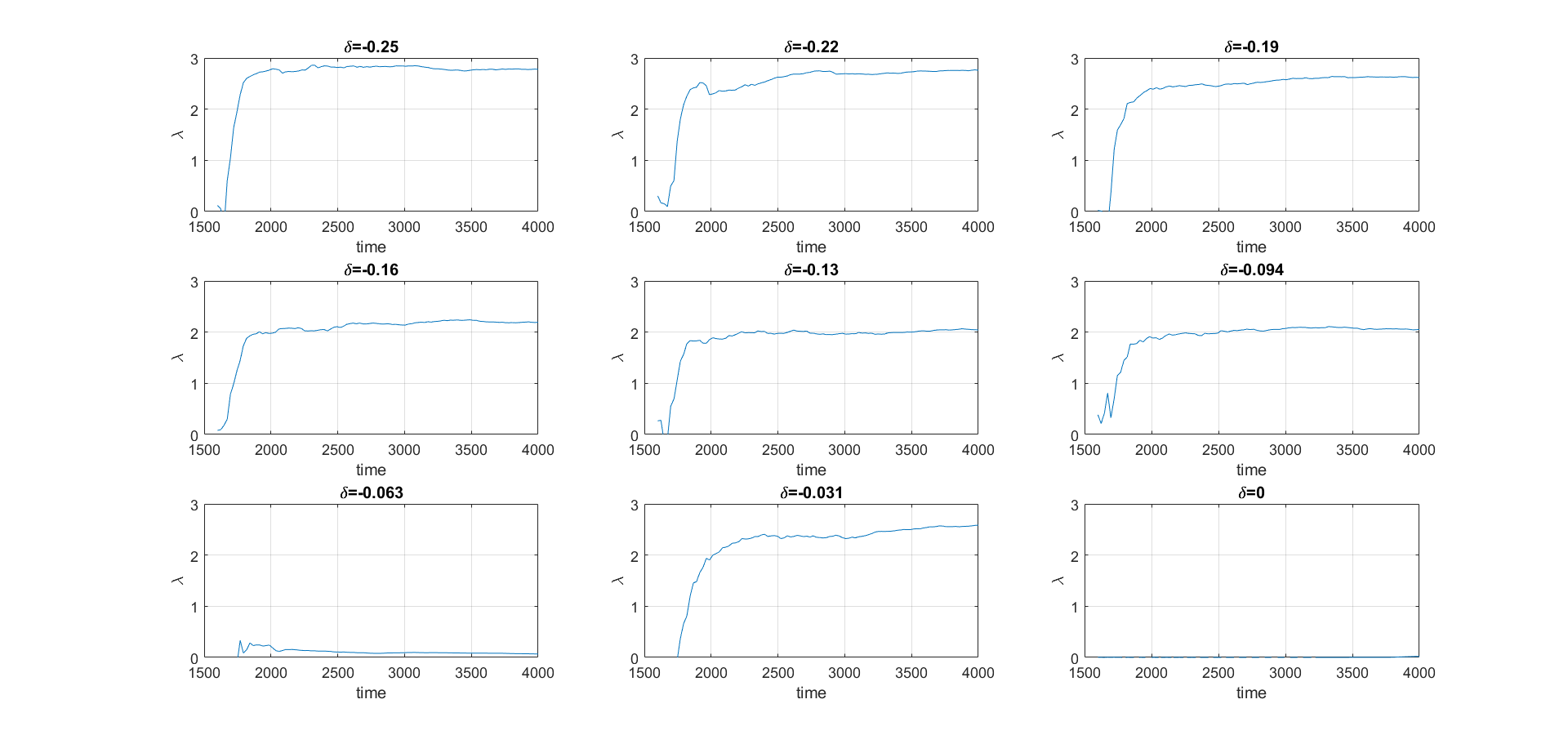}

\caption{Evolution of finite-time Lyapunov exponent for stationary orbits in
Figure 7, as a function of time $s$.}
\end{figure}

\end{document}